\documentclass[aps,prb,twocolumn,superscriptaddress]{revtex4-2}

\usepackage[colorlinks=true,citecolor=blue]{hyperref}
\usepackage{graphicx}
\usepackage{amsmath}
\usepackage{amssymb}
\usepackage{bbold}
\usepackage{braket}
\usepackage{comment}

\DeclareGraphicsExtensions{.png,.jpg,.eps}
\usepackage{xcolor}

\newcommand{\kup}{\mathbf{k}\uparrow}
\newcommand{\kdown}{\mathbf{k}\downarrow}
\begin{document}

\author{Tania Paul}
\affiliation{International Research Centre MagTop, Institute of Physics, Polish Academy of Sciences, Aleja Lotnik\'ow 32/46, PL-02668 Warsaw, Poland}

\author{V. Fern\'andez Becerra}
\affiliation{International Research Centre MagTop, Institute of Physics, Polish Academy of Sciences, Aleja Lotnik\'ow 32/46, PL-02668 Warsaw, Poland}
\author{Timo Hyart}
\affiliation{International Research Centre MagTop, Institute of Physics, Polish Academy of Sciences, Aleja Lotnik\'ow 32/46, PL-02668 Warsaw, Poland}
\affiliation{Department of Applied Physics, Aalto University, 00076 Aalto, Espoo, Finland}

\title{Interplay of quantum spin Hall effect and spontaneous time-reversal symmetry breaking in electron-hole bilayers II: Zero-field topological superconductivity}

\date{\today}
   
\begin{abstract}
It has been proposed that band-inverted electron-hole bilayers support a phase transition from an insulating phase with spontaneously broken time-reversal symmetry to a quantum spin Hall insulator phase as a function of increasing electron and hole densities. Here, we show that in the presence of proximity-induced superconductivity it is possible to realize Majorana zero modes in the time-reversal symmetry broken phase in the absence of magnetic field. We develop an effective low-energy theory for the system in the presence of time-reversal symmetry breaking order parameter to obtain analytically the Majorana zero modes and we find a good agreement between the numerical and analytical results in the limit of weakly broken time-reversal symmetry. We show  that the Majorana zero modes can be detected in superconductor/time-reversal symmetry broken insulator/superconductor Josephson junctions through the measurement of a $4\pi$ Josephson current. Finally, we demonstrate that the Majorana fusion-rule detection is feasible by utilizing the gate voltage dependence of the spontaneous time-reversal symmetry breaking order parameter. 
\end{abstract}

\maketitle

\section{Introduction}

One of the hallmarks of topological insulators is the spin-momentum locking of the surface states \cite{RevKane, RevZhang, Kane-mele, BHZ,Liu08, QSH_HgTe, Du15, Wu18}, which in the presence of induced superconductivity facilitates the realization of Majorana zero modes (MZMs) \cite{Kane_prox, fu2009josephson, Nilsson}. The special property of these quasiparticles, following from their identical creation and annihilation operators, is that they obey non-Abelian braiding statistics \cite{Beenakker-review, Alicea_2012, flensberg2021engineered}, which could be utilized in topological quantum computing \cite{RevMajoforQcomp, Beenakker20}. In one of the theoretically most elegant setups, the MZM appears at the interface between regions where the helical edge modes of a quantum spin Hall (QSH) insulator are gapped by the proximity effects from a superconductor and a ferromagnetic insulator, respectively \cite{fu2009josephson, Nilsson, Beenakker-review, Alicea_2012, flensberg2021engineered}. From the theoretical perspective, the topologically robust single-mode propagation in each direction along the edge makes these systems ideal for the observation of transport signatures of MZMs, such as the zero-bias conductance peak due to resonant Andreev reflection \cite{LawPRL2009, Mi13} and the $4\pi$ Josephson effect \cite{Kitaev2001, fu2009josephson, Beenakker13}. 

Experimentally, unambiguous observation of MZMs in QSH insulators is still missing even though signatures of edge mode superconductivity \cite{yacobyNP2014, PribiagNatNano2015}, Andreev reflection \cite{KnezSC12} and $4\pi$ Josephson effect \cite{Deacon17} have been observed in these systems. One of the problems is that the creation of  hybrid structures of QSH insulators and ferromagnetic insulators has been quite challenging despite of the recent progress in manufacturing hybrid   nanowire-ferromagnetic insulator devices  
\cite{Vaitieknas2020ZerofieldTS}. Moreover, although the gap can be opened also by applying   sufficiently large external magnetic field instead of the utilization of the ferromagnetic insulator, this has detrimental effects on the quality of the superconductors, and therefore this approach is expected to lead to similar difficulties which have so far prevented the unambiguous  observation of the MZMs in the nanowire devices. 

In this paper, we demonstrate that neither the ferromagnetic insulator nor the external magnetic field is needed for the realization of the MZMs. Our approach is based on the previous theoretical work, where it was shown that band-inverted electron-hole bilayers support an unconventional topological phase transition from trivial to the QSH insulator phase via an intermediate insulating phase with spontaneously broken time-reversal symmetry (TRS), arising from the excitonic correlations between the electrons and holes \cite{Pikulin14}. This exotic TRS broken phase is one of the most prominent candidates for the correlated phases appearing in band-inverted semiconductors due to Coulomb interactions \cite{Naveh96, Pikulin14, Budich14, Hu17, Xue18, Zhu19, Varsano20, zeng2021inplane}, and it is consistent with the accumulating experimental evidence of excitonic phenomenology reported in InAs/GaSb quantum wells \cite{Du-exciton, Wu19, Du19PRB, Xiao19, Irie-gap20} as well as in WTe$_2$ \cite{WTe2-exciton, WTe2-exciton2}. Moreover, the properties of the TRS broken phase provide a comprehensive explanation \cite{Paul1} of the temperature, voltage and length dependencies of the observed conductance in InAs/GaSb bilayers \cite{Knez14, Spanton14, Du15, Du-transport}.

We show that MZM appears at the interface between the regions where the helical edge modes are gapped by the TRS breaking order parameter and proximity-induced superconductivity, respectively. Because the TRS is intrinsically broken in this system the MZMs can be realized in the absence of magnetic field and ferromagnetic insulators. We study the Josephson effect in superconductor/time-reversal symmetry broken insulator/superconductor junctions, and conclude, by checking that all the necessary conditions are satisfied \cite{Deacon17, Attila19}, that the MZMs in this system can be detected through the measurement of the  $4\pi$ Josephson effect. We compare various device geometries and vary the tunable parameters of the system to find the optimal conditions for the observation of the $4\pi$ Josephson effect. Finally, building on the previous proposals for the manipulation of MZMs \cite{van_Heck_2012, Hyart13, Heck_2015, Aasen16}, we demonstrate that the Majorana fusion-rule detection is also feasible in this system by utilizing the gate voltage dependence of the spontaneous time-reversal symmetry breaking order parameter.

\section{Phase diagram for zero-field topological superconductivity}

Our starting point is the minimal model for band-inverted electron-hole bilayers \cite{Liu08, liu2013models, Pikulin14, Paul1}
\begin{equation}
H_0 =  \big(E_G - \frac{\hbar^2 k^2}{2m}\big) \tau_z \sigma_0    + A k_x \tau_x \sigma_z   - A k_y \tau_y \sigma_0 + \Delta_{z} \tau_y  \sigma_y,  \label{eq:BHZ}
\end{equation}
where $\tau$'s and $\sigma$'s denote the Pauli matrices in the electron-hole and spin basis, the band-inversion parameter $E_G$ is defined so that for $E_G>0$ ($E_G<0$) the electron and hole bands are (not) inverted at the $\Gamma$ point, $A$ describes the tunneling between layers, $m$ is the effective mass, and $\Delta_z$ is a spin-orbit coupling term  arising due to bulk inversion asymmetry. 
We have ignored the asymmetry of the masses and the momentum-dependent spin-orbit coupling terms, because they are not essential for understanding the phase diagram of the InAs/GaSb bilayers \cite{Pikulin14, Paul1}. The main effect of Coulomb interactions 
is the binding of the electrons and holes into excitons with the characteristic size $d_0$ and binding energy $E_0$ determined by the relation 
$
E_0=\hbar^2/ (m d_0^2)=e^2/(4\pi \epsilon \epsilon_0d_0).
$
This leads to an excitonic mean field \cite{Pikulin14, Paul1}
\begin{eqnarray}
 H_{EC}&=&  \Re[\Delta_1] \tau_y \sigma_y +\Re[\Delta_2] \big[k_x \tau_x \sigma_z  -  k_y \tau_y \sigma_0  \big] \nonumber\\ && + \Im[\Delta_1] \tau_x  \sigma_y -\Im[\Delta_2] \big[k_x \tau_y \sigma_z  + k_y \tau_x \sigma_0   \big],  \label{ec-mean}
\end{eqnarray}
where the gap equations for the $s$-wave and $p$-wave excitonic correlations $\Delta_1$ and $\Delta_2$ are  \cite{Paul1}
\begin{eqnarray}
\Delta_1&=&\frac{g_s d_0^2}{(2\pi)^2} \int d^2 k \ \big[\langle c_{\mathbf{k}\downarrow2}^\dag c_{\mathbf{k}\uparrow 1} \rangle-\langle c_{\mathbf{k}\uparrow2}^\dag c_{\mathbf{k}\downarrow 1} \rangle \big] \nonumber \\
\Delta_2&=&\frac{g_p d_0^4}{(2\pi)^2} \int d^2 k \ \big[ -\langle c_{\mathbf{k}\uparrow2}^\dag c_{\mathbf{k}\uparrow 1} \rangle(k_x-ik_y) \nonumber \\ && \hspace{2.2cm}+\langle c_{\mathbf{k}\downarrow2}^\dag c_{\mathbf{k}\downarrow 1} \rangle (k_x+ik_y) \big]. \label{exc-mean-fields}
 \end{eqnarray}
Here $g_s$ ($g_p$) is the effective interaction strength for $s$-wave ($p$-wave) pairing and $c_{1 \sigma k}$ ($c_{2 \sigma k}$) is the electron annihilation operator  with spin $\sigma$ and momentum $k$ in electron (hole) layer. In our calculations the integration is performed over the range $|\mathbf{k}| \leq 2.26/d_0$, but the exact values of the integration limits are not important.

The values of the model parameters for InAs/GaSb can be estimated by combining theoretical calculations \cite{Liu08, liu2013models, Naveh96, Pikulin14} and the experimentally observed energy gaps \cite{Du15, Du-exciton}. This way, we arrive to parameter values  \cite{Paul1}: $E_0/k_B = 200$ K, $d_0=10$ nm,  $A/(E_0d_0)=0.06$,  $\Delta_z/E_0=0.02$, $g_s/E_0=1.0$ and $g_p/E_0=0.2$. The gate-voltage dependent parameter $E_G$  is varied in our calculations to tune the system from trivial insulator to QSH insulator phase.  For small (large) values of $E_G$ the system is in a trivial (QSH) insulator phase, and these two phases are separated from each other by an insulating phase with spontaneously broken TRS, where $\Im[\Delta_1], \Im[\Delta_2] \ne 0$ \cite{Pikulin14, Paul1}. The bulk gap $\Delta_{\rm bulk}$ remains open for all values of $E_G$, because the intermediate TRS broken phase  enables the connection of the topologically distinct phases without bulk gap closing. The edge gap $\Delta_{\rm edge}$ decreases monotonously when one starts from the trivial phase and tunes the system across the TRS broken phase to the QSH phase, where the gapless edge excitations are protected by the topology \cite{Pikulin14, Paul1}.

Here, we consider the properties of the system in the presence of proximity-induced superconductivity in certain regions of the sample.  The Bogoliubov-de Gennes Hamiltonian in the Nambu basis $\Psi=(c_{\kup},c_{\kdown},c^{\dagger}_{-\kdown},-c^{\dagger}_{-\kup})$ can be written compactly as
\begin{align}
H_{BdG}(\mathbf{k}, \mathbf{x}) =\begin{pmatrix}
H(\mathbf{k}, \mathbf{x}) & \Delta_s(\mathbf{x})\\
\Delta^*_s(\mathbf{x}) & -\sigma_yH^T(-\mathbf{k},  \mathbf{x})\sigma_y
\end{pmatrix},
\label{BdGen}
\end{align}
where 
$H(\mathbf{k}, \mathbf{x})=H_0(\mathbf{k}, \mathbf{x})+H_{EC}(\mathbf{k}, \mathbf{x})$. 
The induced superconductivity has two effects in the regions proximitized by the superconductor. First, it leads to induced superconducting gap $|\Delta_s(\mathbf{x})|=\Delta_0 \ne 0$ in the regions of $\mathbf{x}$ covered by the superconductor. Secondly, it affects the parameters of the normal state Hamiltonian $H(\mathbf{k}, \mathbf{x})$. We assume that the superconductor completely screens the Coulomb interactions and renormalizes the band-inversion parameter, so that $H_{EC}(\mathbf{k}, \mathbf{x})=0$ and $E_G(x)=E_G^{S}$ in the regions of $\mathbf{x}$ covered by superconductors. In the normal regions  the spatially-dependent parameters have values $|\Delta_s(\mathbf{x})|=0$, $E_G(x)=E_G^N$ and $H_{EC}(\mathbf{k}, \mathbf{x})$ is determined by Eqs.~(\ref{ec-mean}) and  (\ref{exc-mean-fields}).

\begin{figure}
    \centering
    \includegraphics[width=0.96\columnwidth]{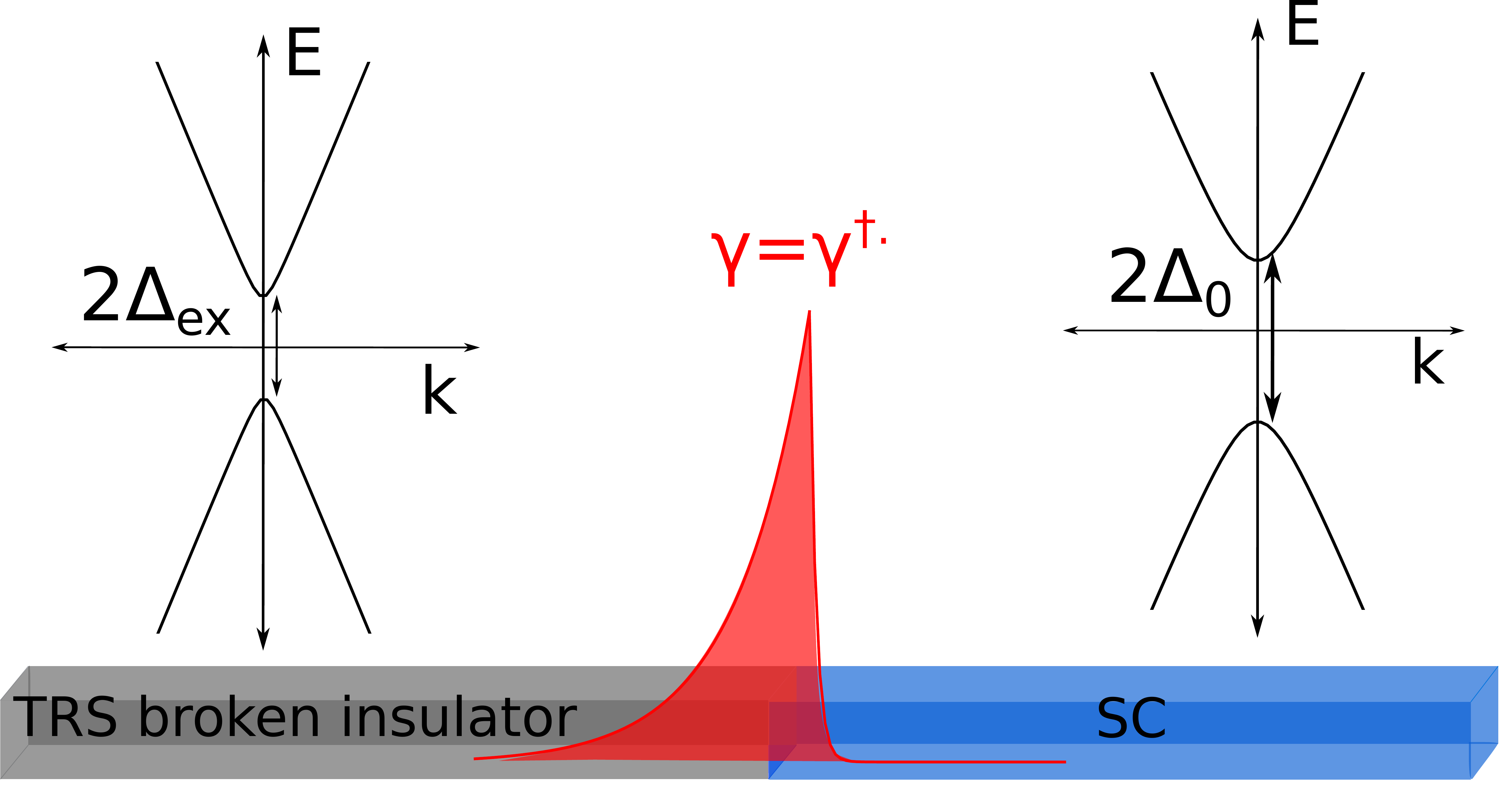}
    \caption{Schematic illustration of MZM $\gamma=\gamma^\dag$ localized at the interface between regions, where the helical edge modes are gapped by the TRS breaking order parameter $\Delta_{{\rm ex}}$ and proximity-induced superconducting pairing amplitude $\Delta_{0}$.}
    \label{fig:single_junction}
\end{figure}

We start by investigating the edge excitations in the presence of induced superconductivity and time-reversal symmetry breaking order parameter. For this purpose we utilize a low-energy theory (valid at energies much smaller than the bulk gap $|E| \ll \Delta_{\rm bulk}$)
\begin{equation}
H_e=\begin{pmatrix}
A_{{\rm eff}}(x)k_x &-i\Delta_{{\rm ex}}(x)  &\Delta_s(x)  & 0\\
i\Delta_{{\rm ex}}(x) & -A_{{\rm eff}}(x)k_x & 0 &\Delta_s(x) \\
\Delta_s^*(x) & 0 &  -A_{{\rm eff}}(x)k_x &-i\Delta_{{\rm ex}}(x)\\
0 & \Delta_s^*(x) & i\Delta_{{\rm ex}}(x) &A_{{\rm eff}}(x)k_x 
\end{pmatrix},
\label{effham}
\end{equation}
where $A_{{\rm eff}}(x)$ is the velocity of the helical edge states, $\Delta_{{\rm ex}}(x)$ is the time-reversal symmetry breaking order parameter in the normal regions of $x$, and $\Delta_s(x)=\Delta_0(x) e^{i \varphi(x)}$ is the induced superconducting pairing potential in the regions of $x$ covered by the superconductors. If the TRS broken insulator (superconductor) covers the region $x<0$ ($x>0$) there  is a MZM localized at $x=0$ (see Fig.~\ref{fig:single_junction}). Namely, there exist a zero-energy solution of the Hamiltonian (\ref{effham}) of the form (see Appendix~\ref{app:analytics})
\begin{equation}
\psi(x)=  \begin{cases} \frac{1}{\cal{N}} \begin{pmatrix}
- e^{-i \pi/4} e^{i\varphi/2}   \\
  e^{-i \pi/4} e^{i \varphi/2}\\
 e^{i \pi/4} e^{-i \varphi/2}\\
 e^{i \pi/4} e^{-i \varphi/2}
\end{pmatrix} e^{\Delta_{{\rm ex}} x/ A^N_{{\rm eff}}}, & x<0 \\
\frac{1}{\cal{N}} \begin{pmatrix}
- e^{-i \pi/4} e^{i\varphi/2}  \\
e^{-i \pi/4} e^{i \varphi/2}\\
e^{i \pi/4} e^{-i \varphi/2}\\
 e^{i \pi/4} e^{-i \varphi/2}  
\end{pmatrix}e^{-\Delta_0 x/A^S_{{\rm eff}}}, & x>0
\end{cases}
\end{equation}
where we can approximate $A^N_{\rm eff}=A+|\Delta_2|$, $A^S_{\rm eff}=A$  and $\Delta_{{\rm ex}}=\sqrt{\Im[\Delta_1]^2+\Im[\Delta_2]^2}$.
The corresponding field operator $\gamma$ in the second quantized form obeys $\gamma=\gamma^\dagger$, and by choosing the normalization constant ${\cal N}$ properly we obtain $\gamma^2=1$. Therefore, this solution satisfies the algebra of the MZMs.

\begin{figure}
  \centering
  \includegraphics[width=\columnwidth]{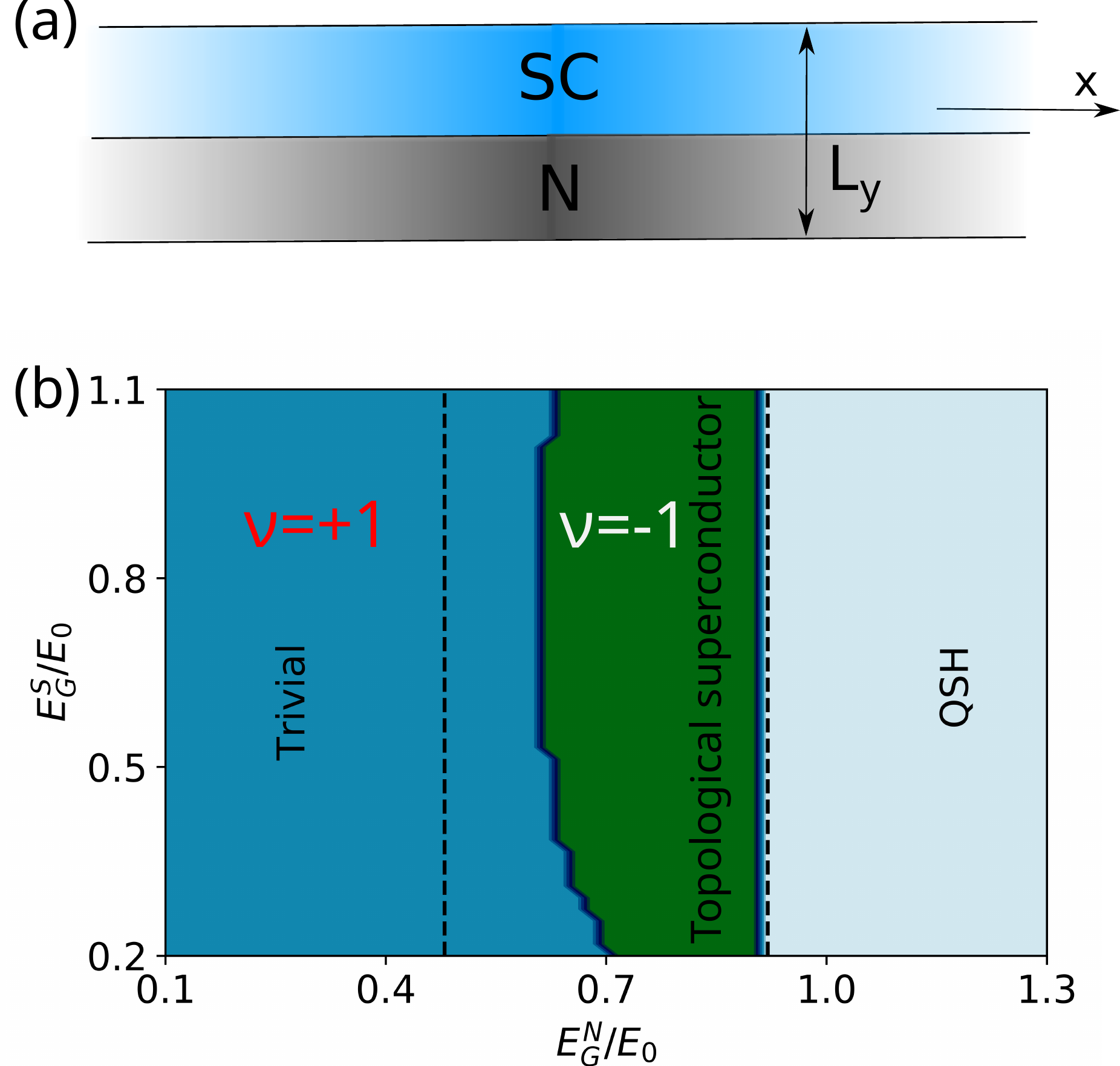}  
  \caption{(a) Topological phase diagram can be determined by computing the $\mathbb{Z}_2$ invariant $\nu$, Eq.~(\ref{Z2-invariant}), for a wide sample $L_y \gg d_0$ with upper half covered by the superconductor and lower half in the normal state. 
  (b) Topological phase diagram as a function of $E_G^S$ and $E_G^N$. In the dark blue trivial phase $\nu=1$ and the hybrid system does not support MZMs. In the green topologically nontrivial phase $\nu=-1$ and the system supports MZMs at the interface of the TRS broken insulator and superconducting regions.  The dashed lines indicate the values of $E_G^N$, where the normal half of the system has transitions from the TRS broken insulator phase to trivial insulator and QSH insulator phases. 
  Here we have used $L_y=100 d_0$ and $\Delta_0=0.1 E_0$.
  } 
   \label{fig:pfaffian_phase}
\end{figure}

We can also study the appearance of the MZM beyond the limits of validity of the effective edge theory using the full two-dimensional Hamiltonian (\ref{BdGen}). For this purpose we consider a sample with region $-L_y/2 \leq y \leq 0$ in the normal state and region $0 \leq y \leq L_y/2$ covered by a superconductor [see Fig.~\ref{fig:pfaffian_phase}(a)]. Such kind of system supports a $\mathbb{Z}_2$ topological invariant \cite{Kitaev2001}
\begin{equation}
    \nu=\mathrm{sgn}[{\rm Pf}M(0) {\rm Pf}M(\pi)], \ M(k_x)=\tau_y\sigma_y H_{BdG}(k_x), \label{Z2-invariant}
\end{equation}
where the Pfaffians of the antisymmetric matrices  $M(k_x=0,\pi)$ are real. The topologically nontrivial (trivial) gapped phases with $\nu=-1$ ($\nu=1$) have odd (even) ground state parity and they do (do not) support unpaired MZMs at the end of the system, which for the geometry shown in Fig.~\ref{fig:pfaffian_phase}(a) corresponds to the interface of the normal and superconducting regions along the edge. Additionally, the system can also support gapless phases.  

We have numerically calculated the topological invariant $\nu$ as a function of $E_G^S$ and $E_G^N$ using the algorithm developed in Ref.~\onlinecite{PfPackage} [see Fig.~\ref{fig:pfaffian_phase}(b)]. For small values of $E_G^N$ the normal region is in a trivial insulator  phase and the hybrid system is also in a topologically trivial phase  $\nu=1$. Upon increasing $E_G^N$ the normal region enters the TRS broken phase [dashed vertical line around $E_G^{N}=0.48 E_0$ in Fig.~\ref{fig:pfaffian_phase}(b)] but at this transition the hybrid system still remains trivial with $\nu=1$. Only by further increasing $E_G^N$  we find a separate phase transition of the hybrid system to a topologically nontrivial phase with $\nu=-1$, and this transition can be controlled with both $E_G^N$ and $E_G^S$ [see Fig.~\ref{fig:pfaffian_phase}(b)]. If we increase $E_G^N$ further the normal region finally enters into the QSH phase so that the hybrid system becomes gapless and the MZMs leak and delocalize into the QSH insulator region along the edge  [see Fig.~\ref{fig:pfaffian_phase}(b)]. We have checked with comprehensive numerical calculations that within the topologically nontrivial phase with $\nu=-1$ the system always supports MZMs at the interfaces of the TRS broken insulator and superconductor regions.  These results are consistent with our effective edge theory, which is expected to the be valid on the QSH side of the TRS broken phase where the edge gap is smaller than the bulk gap, but the numerical approach allows us to establish the phase transition line also on the trivial side of the TRS broken phase.

\section{$4\pi$ Josephson effect}

\begin{figure}[h!]
    \centering
    \includegraphics[width=0.99\linewidth]{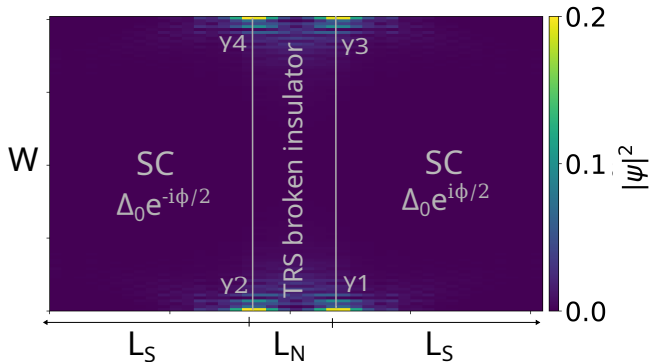}
    \caption{Superconductor/TRS broken insulator/ superconductor Josephson junction for detection of the MZMs via the $4\pi$ Josephson effect. The system supports two MZMs $\gamma_1$ and $\gamma_2$ ($\gamma_3$ and $\gamma_4$) on the bottom (top) edge with the corresponding low-energy local density of states indicated with the colors. The hybridization of the MZMs across a TRS broken regime of length $L_N$ gives rise to a $4\pi$-periodic component in the Josephson current-phase characteristic $I(\phi)$. The width is assumed to be large $W \gg d_0$ so that the parities ${\cal P}_{12}=i\gamma_1 \gamma_2$ and ${\cal P}_{34}=i\gamma_3 \gamma_4$ within each edge are conserved. 
    }
    \label{fig:josephson_junction}
\end{figure}

\begin{figure}
    \centering
   \includegraphics[width=0.9\linewidth]{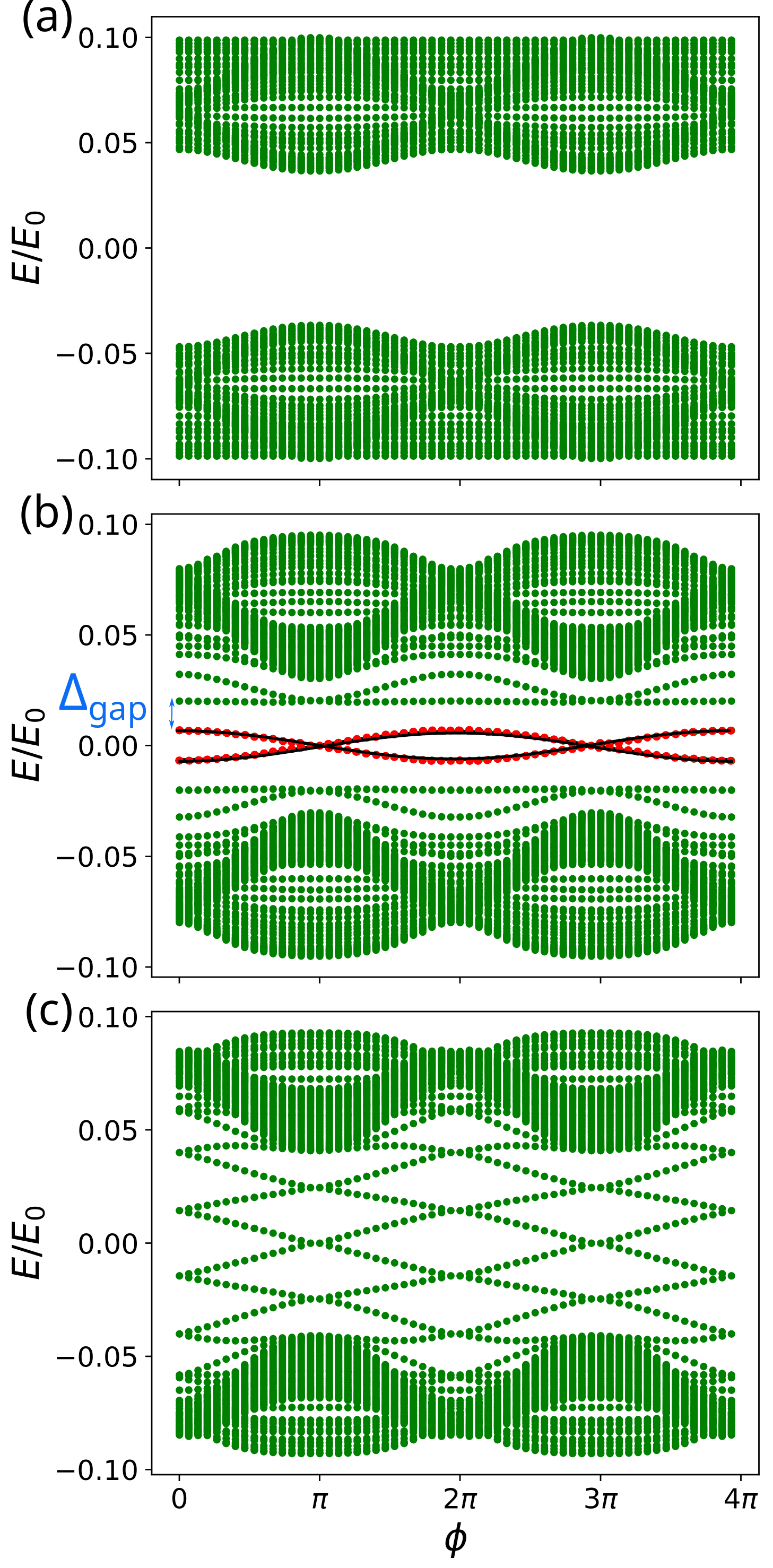}
    \caption{Spectrum of the Josephson junction  (Fig.~\ref{fig:josephson_junction}) as a function of the phase bias $\phi$ when the hybrid system is in (a) $\nu=1$ ($E_G^N=0.3E_0$), (b) $\nu=-1$ ($E_G^N=0.86E_0$) and (c) QSH ($E_G^N=1.1E_0$) part of the phase diagram in Fig.~\ref{fig:pfaffian_phase}.  (a) All the Andreev bound states are gapped. (b) The hybridization of MZMs across the TRS-broken regime leads to $\phi$ dependence of their energies (red lines corresponding to each edge). There exists an energy gap $\Delta_{\rm gap}$ between the MZMs and other Andreev levels. (c) In the QSH regime $\Delta_{\rm gap}=0$.
    In all figures the other parameters are $E_G^S=1.3E_0$, $\Delta_0=0.1E_0$, $L_N=10d_0$, $W=150d_0$ and $L_S=15d_0$.
    }
   \label{fig:spectrum}
\end{figure}

We now proceed to the consideration of the experimental signatures of the MZMs in superconductor/TRS broken insulator/superconductor Josephson junctions. We first consider the simplest geometry, where two large superconducting leads are connected by a wide $W \gg d_0$ normal region of length $L_N$  as shown in  Fig.~\ref{fig:josephson_junction}. Such a Josephson junction has a qualitatively different spectrum (see Fig.~\ref{fig:spectrum}) depending on whether the hybrid system is in a $\nu=1$, $\nu=-1$ or QSH part of the phase diagram shown in Fig.~\ref{fig:pfaffian_phase}. In the trivial phase $\nu=1$ there exists only gapped Andreev bound states [Fig.~\ref{fig:spectrum}(a)], and therefore the application of a phase bias across the superconductors 
 $\Delta_s(\mathbf{x})=\Delta_0\bigl(e^{i\phi/2}\theta(x-L_n)+e^{-i\phi/2}\theta(-x)\bigr)$ leads to a conventional $2\pi$ periodic Josephson effect. In the topologically nontrivial $\nu=-1$ state, the system supports four MZMs $\gamma_i$ ($i=1,...,4$) at the interfaces of the TRS broken insulator and superconducting regions [see Fig.~\ref{fig:josephson_junction} and Fig.~\ref{fig:spectrum}(b)]. Because $W \gg d_0$ the coupling between the Majoranas across the width of the sample can be neglected so that the parities ${\cal P}_{12}=i\gamma_1 \gamma_2$ and ${\cal P}_{34}=i\gamma_3 \gamma_4$ within each edge are good quantum numbers, and the application of the phase bias $\phi$  leads to a Josephson current, which in addition to the conventional $2\pi$-periodic component also has parity-dependent $4\pi$-periodic component \cite{Kitaev2001, fu2009josephson, Beenakker13} 
 \begin{equation}
 I(\phi)=I_{2 \pi} \sin (\phi) + I_{4 \pi} \frac{{\cal P}_{12}+{\cal P}_{34}}{2} \sin (\phi/2)+h.h.,
 \end{equation}
where h.h. denotes the higher harmonics. The magnitude of the $4\pi$ Josephson effect 
\begin{equation}
I_{4\pi}=\frac{e}{\hbar} \frac{2}{4\pi}\int_0^{4\pi} d\phi   \sum_M\frac{dE_M(\phi)}{d\phi} \sin(\phi/2)
\end{equation}
is determined by the quasiparticle energies $E_M(\phi)$, originating from the hybrization of MZMs $\gamma_1$ and $\gamma_2$ ($\gamma_3$ and $\gamma_4$) at bottom (top) edge, which cross $E=0$ at $\phi=\pi$ [doubly degenerate red lines in Fig.~\ref{fig:spectrum}(b)]. In the asymptotic limit $L_N \gg A^N_{{\rm eff}}/\Delta_{{\rm ex}}$ (see Appendices \ref{app:analytics} and \ref{app:Josephson})
\begin{equation}
 I_{4\pi}= 2  \frac{e}{\hbar}\frac{\Delta_0 \Delta_{{\rm ex}}}{\Delta_0+\Delta_{{\rm ex}}} e^{-\Delta_{{\rm ex}} L_N/A^N_{{\rm eff}}}. \label{I_4pi-analytic_app}
 \end{equation}
On the other hand,  
\begin{equation}
I_{2\pi}= \frac{e}{\hbar} \frac{2}{2\pi}\int_0^{2\pi} d\phi   \sum_k\frac{dE_k(\phi)}{d\phi} \sin\phi
\end{equation}
includes contributions from all other Andreev levels $k$ with $E_k<0$ except the MZMs.
The methods for calculating $I(\phi)$, including hybrid Kernel polynomial method \cite{irfan2019hybrid}, exact diagonalization and the low-energy effective edge theory, are discussed in Appendix \ref{app:Josephson}. In all our calculations the current is expressed in units of $I_0=eE_0/\hbar \approx 4 \ \mu$A. We concentrate only on the $I_{2 \pi}$ and $I_{4 \pi}$, which can be experimentally measured independently from the other harmonics in the Josephson radiation spectrum   \cite{Deacon17, Attila19}. In this type of experiments, the applied voltage $V$ across the Josephson junction leads to ac Josephson effect $\phi(t)=2eV t/\hbar$, so that $I_{4 \pi}$ ($I_{2 \pi}$) results in Josephson radiation at frequency $f_{4 \pi}=eV/h$ ($f_{2 \pi}=2eV/h$). In addition to $I_{4 \pi}$ and  $I_{2 \pi}$ also the energy gap $\Delta_{\rm gap}$ between the MZMs and other Andreev levels is important for the robustness of the $4\pi$ Josephson effect.  In the presence of TRS breaking order parameter $\Delta_{\rm gap} \ne 0$, whereas in the QSH regime $\Delta_{\rm gap}=0$ [cf.~Fig.~\ref{fig:spectrum}(b),(c)].

Several conditions need to be satisfied so that the $4\pi$ Josephson effect can be robustly detected. (i) It is important that the frequencies are much larger than the quasiparticle poisoning rate and the hybridization of the MZMs localized at the different edges $f_{4 \pi} \gg 1/T_{\rm pois}, \Delta_{\rm hyb}/h$. These do not pose fundamental problems, because the hybridization $\Delta_{\rm hyb}$ can be made arbitrarily small by increasing the width of the sample $W$ and the quasiparticle poisoning time $T_{\rm pois}$ can be as large as seconds in the state-of-the-art superconducting devices in the absence of magnetic field \cite{Dima18,mannila2021superconductor,Dima21}. Nevertheless, in practice the experiment is expected to be  less challenging if the Josephson radiation frequencies are reasonably large. Indeed, based on the earlier experiments \cite{Deacon17, Attila19}, we expect that the ideal operation regime for probing the $4\pi$ Josephson effect is achieved by tuning the Josephson frequencies to the gigahertz frequency range with the applied voltage $V$.   (ii)  The energy gap $\Delta_{\rm gap}$ between the MZMs and the other Andreev bound states  should be sufficiently large to avoid Landau-Zener tunneling $\hbar d\phi/dt=h f_{2\pi} \ll \Delta_{\rm gap}$ and thermal excitations $k_BT \ll \Delta_{\rm gap}$. Violation of these conditions gives rise to $2 \pi$-periodic occupation of the Andreev levels which suppresses the $4\pi$ Josephson effect, and the remaining signatures of $4\pi$ Josephson effect depend on the details of the relaxation processes \cite{Beenakker13, Sticlet18, Deacon17}. Thus, in the most robust operation conditions $\Delta_{\rm gap}/h > 10$ GHz.  (iii) Finally, $I_{4\pi}$ should be sufficiently large to overcome the detector sensitivity. Based on the earlier experiment \cite{Deacon17,Attila19}, we estimate that  $I_{4\pi}>30$ nA  would allow robust detection of the $4\pi$ Josephson effect. Although $I_{4\pi}$ and $I_{2\pi}$ give separate peaks in the ideal detection, in practice very large $I_{2\pi}$ can cause problems due to the broadening of the peaks in the case of imperfect detection of the radiation spectrum.

\begin{figure}
    \centering
    \includegraphics[width=\linewidth]{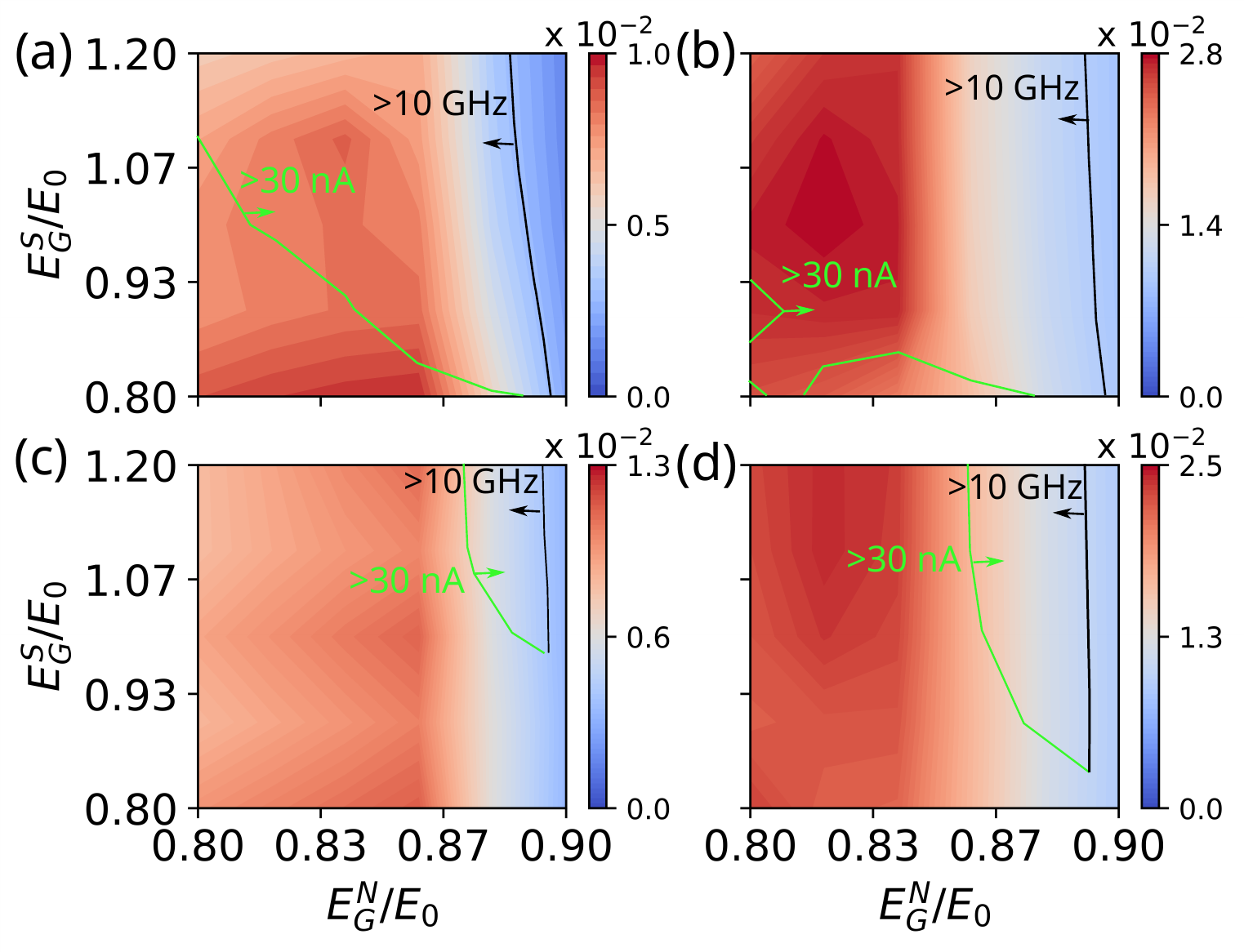}
    \caption{Energy gap $\Delta_{\rm gap}/E_0$ between the MZMs and the other Andreev levels as a function of $E_G^N$ and $E_G^S$ for (a) $\Delta_0=0.03E_0$ and $L_N=5 d_0$, (b) $\Delta_0=0.1 E_0$ and $L_N=5 d_0$, (c) $\Delta_0=0.03 E_0$ and $L_N=10 d_0$, and (d) $\Delta_0=0.1 E_0$ and $L_N=10 d_0$. In all figures $L_S=15d_0$ and $W=150 d_0$.
    The optimal parameter regimes for the observation of the $4\pi$ Josephson effect is the region between black and green lines. The upper bound of $E_G^N$ (black line) is determined by the condition that $\Delta_{\rm gap}$ is sufficiently large, whereas the lower bounds of $E_G^N$ and $E_G^S$ (green line) are determined by the condition that $I_{4\pi}$ (shown in Fig.~\ref{fig:contour-4pi}) is sufficiently large.}
    \label{fig:gap}
\end{figure}
\begin{figure}
    \centering
    \includegraphics[width=\linewidth]{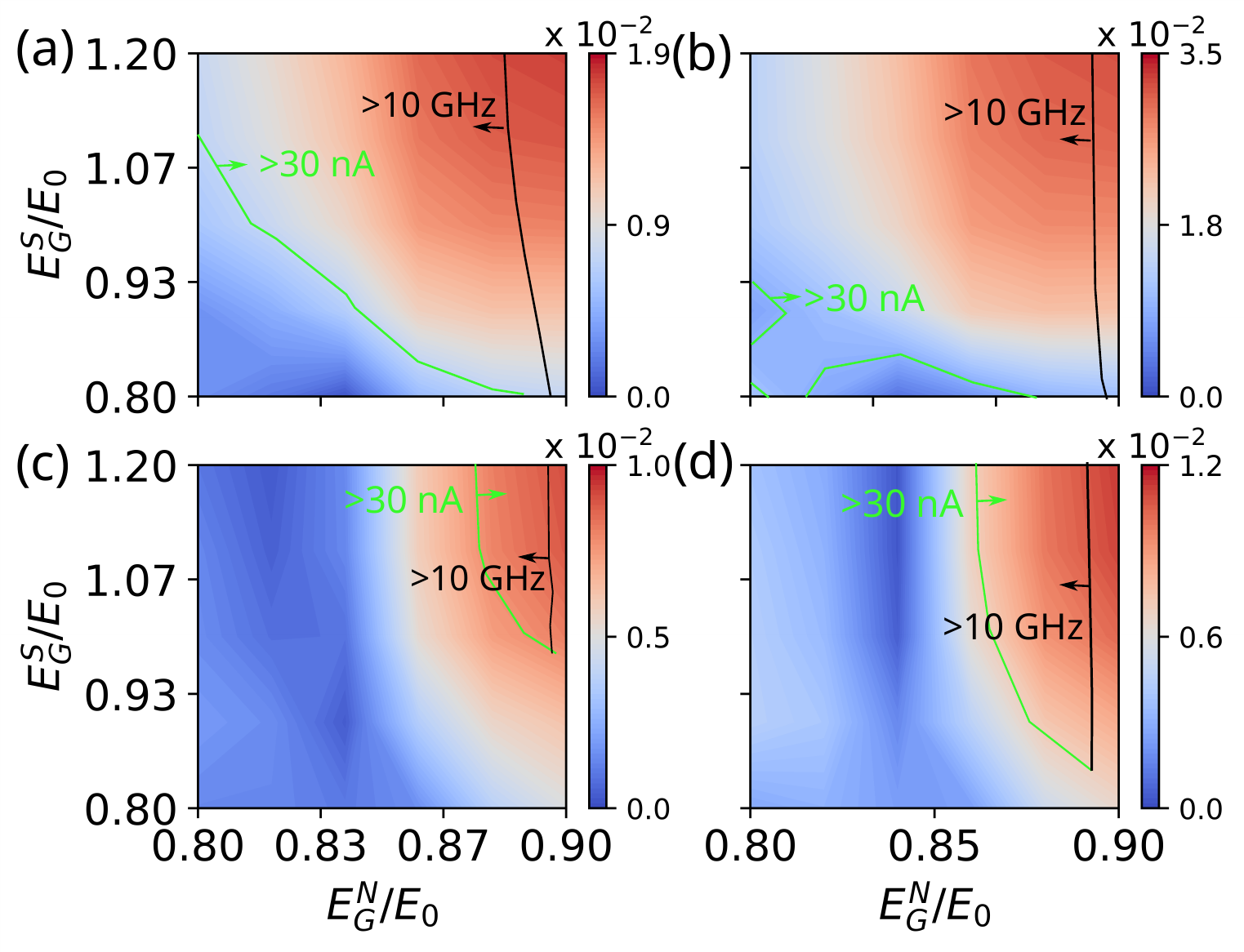}
    \caption{ (a)-(d)  Magnitude of the $4\pi$ Josephson effect $I_{4\pi}/I_0$ for the same set of model parameters as described in Figs.~\ref{fig:gap}(a)-(d), respectively. The requirement that $I_{4\pi}$ is sufficiently large determines the lower bounds of $E_G^N$ and $E_G^S$ in the optimal parameter regimes shown in Figs.~\ref{fig:gap} and \ref{fig:contour-4pi}.}
    \label{fig:contour-4pi}
\end{figure}
\begin{figure}
    \centering
    \includegraphics[width=\linewidth]{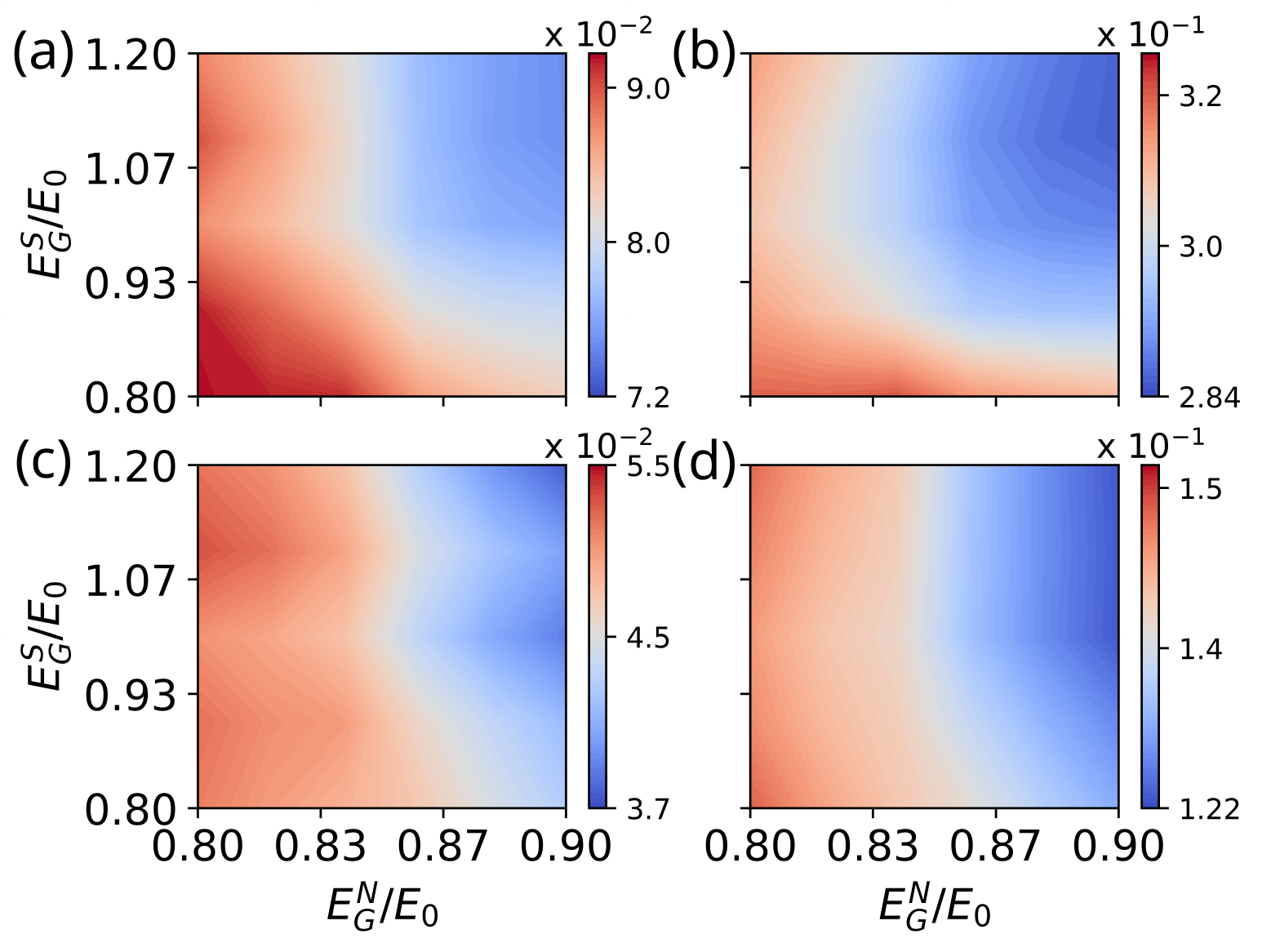}
    \caption{(a)-(d) Magnitude of the $2\pi$ Josephson effect $I_{2\pi}/I_0$ for the same set of model parameters as described in Figs.~\ref{fig:gap}(a)-(d), respectively. $I_{2\pi}$ is typically an order of magnitude larger than $I_{4\pi}$ shown in Figs.~\ref{fig:contour-4pi}. 
    }
    \label{fig:contour-2pi}
\end{figure}

As discussed in the previous Section, the low-energy physics is quite well captured by the effective edge theory. Therefore, $I_{4\pi}$ and $\Delta_{\rm gap}$ can be accurately calculated analytically using  Eq.~(\ref{I_4pi-analytic_app}) and the approach discussed in Appendix \ref{app:analytics} in a large part of the parameter space $E_G^N$, $E_G^S$, $\Delta_0$ and $L_N$ (see Appendix \ref{app:Josephson}). On the other hand, we find that there exists a large number of Andreev levels which contribute to the $I_{2 \pi}$ because they have a significant dispersion as a function of $\phi$ (see Fig.~\ref{fig:spectrum}). Therefore, we utilize the hybrid kernel polynomial method \cite{irfan2019hybrid} implemented within Kwant software package \cite{Groth_2014} for calculation of the $I_{2\pi}$ (see Appendix \ref{app:Josephson}). Our results for $\Delta_{\rm gap}$, $I_{4\pi}$ and $I_{2\pi}$ for various different model parameters are shown in Figs.~\ref{fig:gap}, \ref{fig:contour-4pi} and \ref{fig:contour-2pi}, respectively. We find that there exists a large region in parameter space where the conditions (ii) and (iii) for the robust detection $4\pi$ Josephson effect are satisfied (see the optimal parameter regimes indicated in Figs.~\ref{fig:gap} and \ref{fig:contour-4pi}).   However, we also notice that $I_{2\pi}$ is typically an order of magnitude larger than $I_{4\pi}$ (c.f.~Figs.~\ref{fig:contour-4pi} and \ref{fig:contour-2pi}).
This is not a problem in the case of an ideal detection because the $4\pi$ and $2\pi$ Josephson effects give rise to peaks at separate frequencies in the Josephson radiation spectrum. However, the large peak caused by $I_{2\pi}$ may overshadow the peak caused by the $I_{4\pi}$ due to the broadening of the peaks in the case of imperfect detection of the radiation spectrum. We point out that if this becomes a problem it is possible to significantly reduce $I_{2\pi}$ by changing the device geometry so that the superconducting leads are coupled only to one of the edges of the sample. Moreover, $I_{2\pi}$ can be reduced also by decreasing the width of the sample $W$ while still keeping it large enough so that coupling between MZMs located at the top and bottom edges remains sufficiently small  (see Appendix \ref{app:Josephson}).

\section{Fusion-rule detection}

\begin{figure}[h!]
    \centering
    \includegraphics[width=\linewidth]{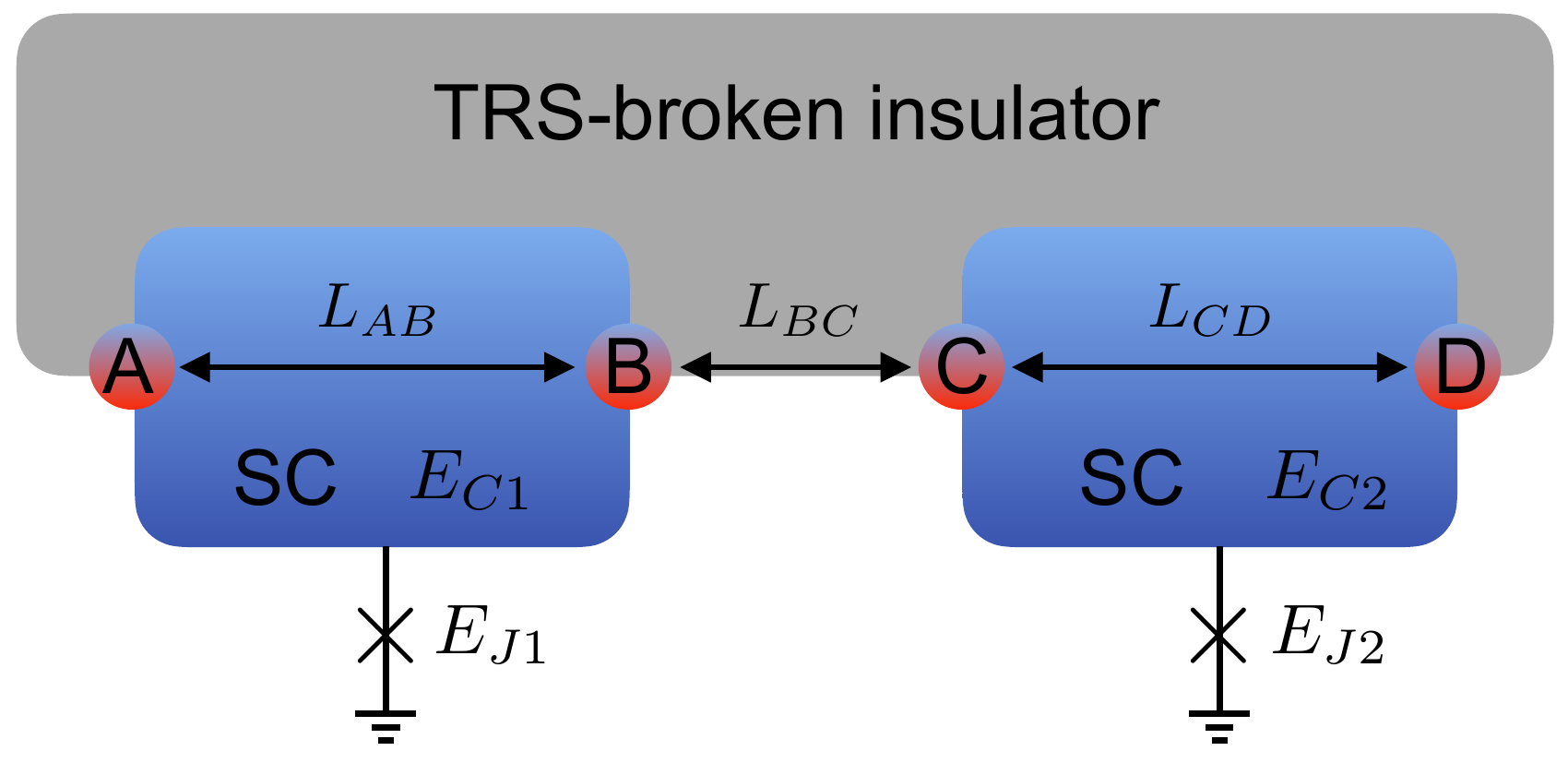}
    \caption{Setup for detection of fusion rules. The couplings between MZMs can be controlled with the tunable Josephson energies $E_{J1}$,  $E_{J2}$, and the gate-tunable energy gap  $\Delta_{\rm ex}$ appearing due to the spontaneous TRS breaking.}
    \label{fig:fusion_setup}
\end{figure}

The fusion rules are a fundamental property of non-Abelian anyons describing how they coalesce. Equivalently with the nontrivial braiding statistics, also the nontrivial fusion rules require the existence of the topological ground state degeneracy, and they can be used to define the non-Abelian anyons, but they are much simpler to detect \cite{Aasen16}. Namely, the
previously considered setups can be easily generalized for the fusion-rule detection and validation of the topological qubit by introducing tunable  Josephson energies $E_{J1}$ and $E_{J2}$ of two superconducting islands coupled to a superconducting ground (see Fig.~\ref{fig:fusion_setup}), in analogy to the proposal suggested in Ref.~\cite{Aasen16} for the fusion-rule detection in  nanowires. The superconducting islands contain a macroscopic number of electrons but they are sufficiently small so that the charging energies of the islands, $E_{C1}$ and $E_{C2}$, exceed temperature. The setup shown in Fig.~\ref{fig:fusion_setup} contains four MZMs, which we denote as $\gamma_{i}$ ($i=A, B, C, D$), and they are spatially separated by distances $L_{AB}$, $L_{BC}$ and $L_{CD}$, respectively. The MZMs satisfy the algebra $\gamma_i=\gamma_i^\dag$ and $\{\gamma_i, \gamma_j \}=2 \delta_{ij}$, and they can be used to define a topological qubit. Assuming that the total parity is fixed (all operations need to performed faster than the quasiparticle poisoning time), the MZMs admit two ground states. For even total parity, we can write these two logical states of the topological qubit as
\begin{equation}
|0_{AB}, 0_{CD} \rangle, \quad  |1_{AB}, 1_{CD} \rangle, \label{basis}
\end{equation}
where $N_{AB}, N_{CD} \in \{0,1\}$  refer to occupation numbers of ordinary fermions $f^\dag_{AB}=(\gamma_A+i \gamma_B)/2$ and $f^\dag_{CD}=(\gamma_C+i \gamma_D)/2$. The measurement of the state in this basis can be performed by coupling (fusing) the MZMs $\gamma_A$ and $\gamma_B$ (or $\gamma_C$ and $\gamma_D$) and then utilizing microwaves, charge sensor or charge pumping for read out of the fermion occupation number as described in detail in Refs.~\cite{Hyart13, Aasen16}. The fusion rule of MZMs is 
\begin{equation}
\sigma \times \sigma = I+\psi, \label{fusion_rule_MZM}
\end{equation}
which states that two MZMs can coalesce into identity $I$ (occupation number of the fermion corresponding to the  pair of MZMs is $0$) or a fermion $\psi$ (occupation number is $1$). The simplest manifestation of this fusion rule in our setup is to initialize the system into a state with  well-defined occupation numbers of the fermions $f^\dag_{AD}=(\gamma_A+i \gamma_D)/2$ and $f^\dag_{BC}=(\gamma_B+i \gamma_C)/2$, and to measure the occupation number $N_{AB}$ (or $N_{CD}$). The probabilities for measuring occupation numbers $0$ and $1$ are then equal, because (see Appendix \ref{app:fusion})
\begin{eqnarray}
 |0_{AD}0_{BC}\rangle&=&\frac 1{\sqrt 2} \big(|0_{AB}0_{CD}\rangle+|1_{AB}1_{CD}\rangle\big), \nonumber \\
  |1_{AD}1_{BC}\rangle&=&\frac i{\sqrt 2} \big(|1_{AB}1_{CD}\rangle-|0_{AB}0_{CD}\rangle\big). \label{fusion-basis}
\end{eqnarray}
However, some care is required to make sure that the outcome of the measurement can be interpreted as evidence of the Majorana fusion rules. We will describe the necessary protocol for the fusion-rule detection \cite{Aasen16} and identify the optimal operation regime below. We point out that the detection of the braiding statistics and more complicated manipulations of the MZMs are also possible using our platform, but these operations would require a branched geometry \cite{Hyart13, Aasen16, van_Heck_2012, Mi13, Heck_2015}, which is much more challenging to realize experimentally. Thus, we do not consider these possibilities in this work.  

The fusion-rule detection protocol outlined in Ref.~\cite{Aasen16} requires that the couplings between the MZMs can be varied as a function of time.
In the limit $E_{Jk} \gg E_{Ck}$ ($k=1,2$), $L_{AB}, L_{CD} \gg A^S_{\rm eff}/\Delta_0$ and $L_{BC} \gg A^N_{\rm eff}/\Delta_{\rm ex}$, the low-energy Hamiltonian for the MZMs is \cite{Hyart13, Aasen16, van_Heck_2012, Mi13, Heck_2015}
\begin{equation}
H_{\rm eff}=-i U_{AB} \gamma_A \gamma_B - i U_{BC} \gamma_B \gamma_C -i U_{CD} \gamma_C \gamma_D, \label{TQ-Ham}
\end{equation}
where  (see Appendix \ref{app:analytics}) 
\begin{widetext}
\begin{equation}
U_{AB (CD)}=\begin{cases}
\frac{16}{(2 \pi^2)^{1/4}} E_{C1(2)} \bigg(\frac{E_{J1(2)}}{E_{C1(2)}}\bigg)^{3/4} e^{-\sqrt{8E_{J1(2)}/E_{C1(2)}}} \cos(\pi q_k/e), & {\rm charging-energy \  dominant} \\
 \frac{\Delta_0 \Delta_{\rm ex}}{\Delta_0+\Delta_{\rm ex}} e^{-\Delta_0 L_{AB (CD)}/A^S_{\rm eff}}, & {\rm Majorana-overlap \  dominant}
\end{cases}
\end{equation}
\end{widetext}
and
\begin{equation}
U_{\rm BC}= \frac{\Delta_0 \Delta_{\rm ex}}{\Delta_0+\Delta_{\rm ex}} e^{-\Delta_{\rm ex} L_{BC}/A^N_{\rm eff}} \cos(\phi/2).
\end{equation}
Here the offset charge $q_k$ can be controlled with the help of a voltage applied to a nearby gate electrode, and we have assumed that $U_{ij} < \Delta_{\rm gap}$, so that the excitations above the energy gap $\Delta_{\rm gap}$ can be neglected.  In the  fusion-rule detection protocol \cite{Aasen16}, it is important that each coupling $U_{ij}$ ($ij=AB, BC, CD)$ can be turned on $U_{ij}=U_{\rm max}$ and off $U_{ij}=U_{\rm min}$, so that $U_{\rm min} \ll U_{\rm max}$. In the case $U_{AB(CD)}$ we assume that $L_{AB (CD)}$ is sufficiently large so that we are always in the charging-energy dominant regime (see conditions below). Thus, these couplings are controlled with $E_{J1(2)}$. Importantly, $U_{AB(CD)}$ depends exponentially on $E_{J1(2)}$, so that a moderate tuning of the Josephson couplings
leads to $U_{\rm min} \ll U_{\rm max}$. We point out that it is only important to be in the regime $E_{J1(2)} \gg E_{C1(2)}$ when the couplings are turned off  $U_{ij}=U_{\rm min}$. When the couplings are turned on $U_{ij}=U_{\rm max}$ the charging energy can even be larger than the Josephson coupling as long as $U_{\rm max} < \Delta_{\rm gap}$ is satisfied. If $E_{J1(2)} \ll E_{C1(2)}$ one obtains $U_{\rm max} = E_{C1(2)}/2$ \cite{Hyart13, Aasen16}.
Similarly, one also needs to tune $U_{BC}$ during the fusion protocol. Because of the exponential dependence on $\Delta_{\rm ex}$, the best way to turn $U_{BC}$ on and off is to control $\Delta_{\rm ex}$ with the gate voltages. Also in this case it is important to be in the asymptotic limit ($\Delta_{\rm ex} \gg A^N_{\rm eff}/L_{BC}$) when the coupling is turned off $U_{BC}=U_{\rm min}$, and it is only important that  $U_{\rm max} < \Delta_{\rm gap}$ is satisfied when the coupling is turned on.  In the description of the fusion rule detection protocol we set $U_{\rm min} \to 0$. In practice it means that the time scale operations $T_{\rm op}$ must satisfy $T_{\rm op} \ll \hbar/U_{\rm min}$. All the operations should be performed adiabatically with respect to the gap $T_{\rm op} \gg \hbar/\Delta_{\rm gap}$ and the  turned-on coupling $T_{\rm op} \gg \hbar/U_{\rm max}$.

\begin{figure}[h!]
    \centering
    \includegraphics[width=0.95\linewidth]{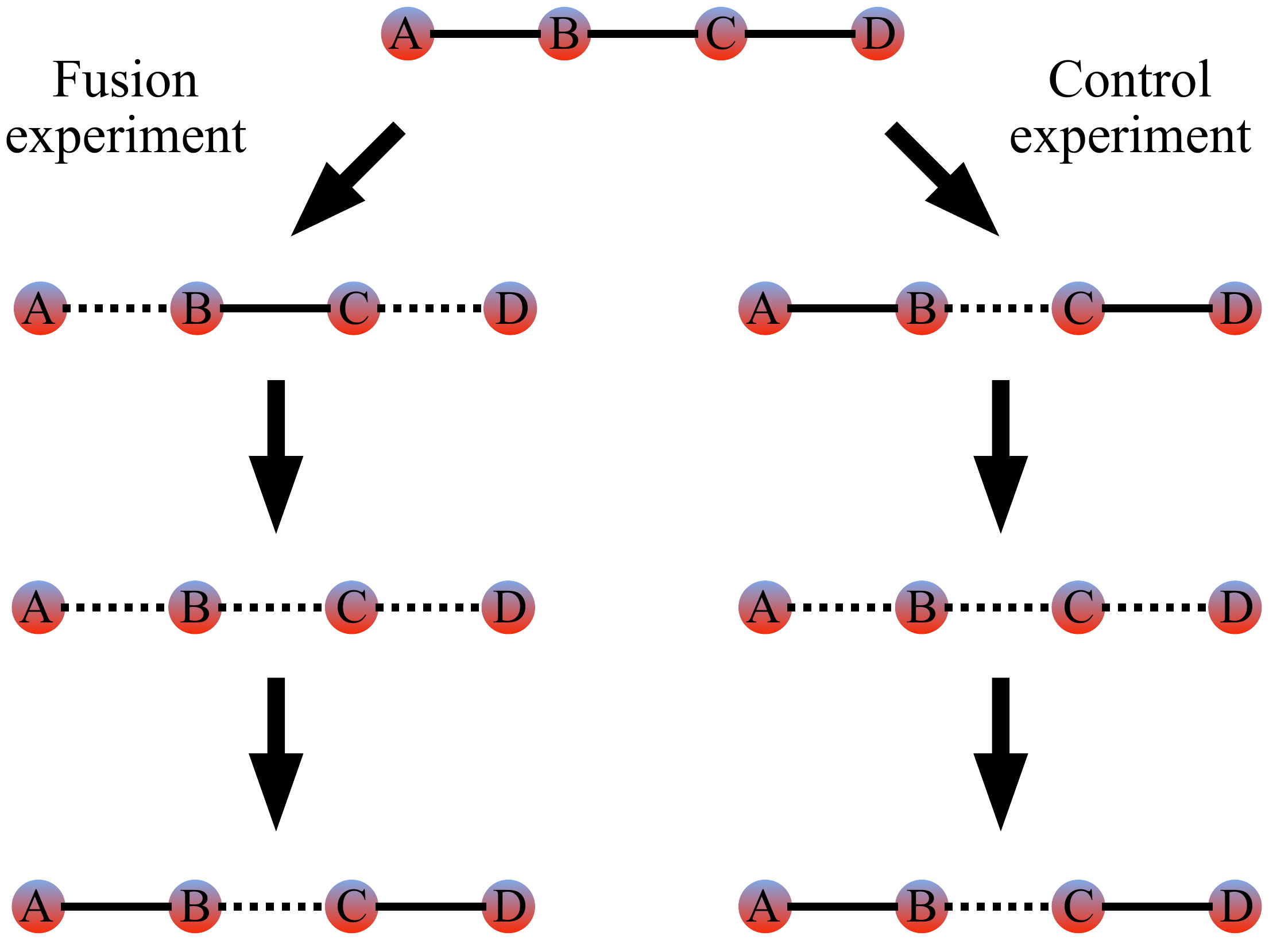}
    \caption{Protocol for detection of fusion rules. The solid (dashed) lines indicate couplings $U_{ij}$ between the MZMs $\gamma_i$ and $\gamma_j$, which are turned on (off). In the final state the occupation numbers $[N_{AB}, N_{CD}]$ are measured. In the fusion experiment the sequence of the operations lead to measurements of  occupation numbers $[0_{AB}, 0_{CD}]$ and $[1_{AB}, 1_{CD}]$ with equal probabilities, reflecting the fusion of MZMs into $I$ and $\psi$ channels. In the control experiment the sequence of operations always leads to fusion of MZMs into $I$ channel so that the measured  occupation numbers are always $[0_{AB}, 0_{CD}]$.}
    \label{fig:fusion_protocol}
\end{figure}

The fusion-rule detection protocol, which is based on the proposal in Ref.~\cite{Aasen16}, is shown in Fig.~\ref{fig:fusion_protocol}. It consists of a two different sequences of operations called fusion experiment and control experiment. In the basis of logical states of the topological qubit (\ref{basis}) the low-energy Hamiltonian (\ref{TQ-Ham}) can be written as 
\begin{equation}
H_{\rm eff}= - (U_{AB}+U_{CD}) \sigma_z - U_{BC} \sigma_x, \label{Ham-log-basis}
\end{equation}
so that the eigenenergies and eigenstates of the Hamiltonian (\ref{Ham-log-basis}) are (we assume $U_{BC}>0$ for simplicity)
\begin{equation}
E_{\pm}=\pm \sqrt{(U_{AB}+U_{CD})^2+U_{BC}^2}, 
\end{equation}
\begin{equation}
\psi_{\pm}=\frac{1}{\sqrt{2}} \bigg( \sqrt{1\mp\frac{U_{AB}+U_{CD}}{|E_{\pm}|}},\mp \sqrt{1\pm\frac{U_{AB}+U_{CD}}{|E_{\pm}|}} \bigg)^T.
\end{equation} 
In both sequences the system is first initialized to a unique ground state of the system $|\psi_{\rm GS} \rangle$, which is determined by the relative magnitudes of the turned-on couplings $U_{AB}$, $U_{BC}$ and $U_{CD}$, but the exact ground state is not important in the following. The fusion and control sequences applied to the ground state $|\psi_{\rm GS} \rangle$  look very similar but they should lead to very different measurement outcomes of the occupation numbers $N_{AB}$ and $N_{CD}$ in the last step of the experiments, allowing to obtain strong evidence of the Majorana fusion rules.

In the fusion experiment one first turns off the couplings $U_{AB}$ and $U_{CD}$, so that $|\psi_{\rm GS} \rangle$ evolves to a state $|\psi_F\rangle=\frac{1}{\sqrt{2}} (|0_{AB}, 0_{CD}\rangle+|1_{AB}, 1_{CD}\rangle)$. This corresponds to the initialization of the system into a state with  well-defined occupation numbers of the fermions $f^\dag_{AD}$ and $f^\dag_{BC}$ as discussed above [see Eq.~(\ref{fusion-basis})]. After this one can turn off the coupling $U_{BC}$ without changing the state of the system and finally measure the occupation numbers $N_{AB}$ and $N_{CD}$. Thus, in the fusion experiment the occupation numbers $[0_{AB}, 0_{CD}]$ and $[1_{AB}, 1_{CD}]$ are measured with equal probabilities, reflecting the fusion of MZMs into $I$ and $\psi$ channels [see Eq.~(\ref{fusion_rule_MZM})]. 

In the control experiment one first turns off the coupling $U_{BC}$. This initializes the system into a state with  well-defined occupation numbers of the fermions $f^\dag_{AB}$ and $f^\dag_{CD}$, e.g.~assuming that $U_{AB}, U_{CD}>0$ we obtain $|\psi_C\rangle=|0_{AB}, 0_{CD}\rangle$. After this one can turn off the coupling $U_{AB}$ and $U_{CD}$ without changing the state of the system and finally measure the occupation numbers $N_{AB}$ and $N_{CD}$. Thus, in the control experiment the occupation numbers $[0_{AB}, 0_{CD}]$  are always measured, reflecting the fusion of MZMs into $I$ channel.

The requirements for the fusion-rule detection mentioned above can be satisfied by choosing the charging energy to be $E_{C1(2)} \sim 0.1$ meV and 
tuning the Josephson energies in the range $E_{J1(2)} \in [0, 50  E_{C1(2)}]$ so that $U_{\rm max} \sim 0.5 E_c$ to $U_{\rm min} \sim 10^{-7} E_c$. Taking $A^{S(N)}_{\rm eff} \sim 0.1 E_0 d_0$ and $\Delta_{\rm ex} \sim \Delta_0 \sim 0.03 E_0$, we find that the Majorana overlap across the superconducting island 1 (2) can be safely neglected if $L_{AB (CD)} \gtrsim 1$ $\mu m$. According to our calculations the gap $\Delta_{\rm ex}$ opened due to TRS symmetry breaking can be tuned with gate voltages to be between $[0.005 E_0, 0.03 E_0]$. Thus, by choosing $L_{BC} \sim 0.5$ $\mu m$, we find $U_{\rm max} \sim 10^{-3}E_0$ and $U_{\rm min} \sim 10^{-8}E_0$. With these values of the parameters also $U_{\rm max} < \Delta_{\rm gap}$, so that we can neglect the bulk excitations in our low-energy theory.
At the same time $k_BT \ll  \Delta_{\rm gap}$ and $k_BT \ll U_{\rm max}$, so that also the thermal excitations can be neglected. The operation time should satisfy $T_{\rm op} \gg  \hbar/\Delta_{\rm gap} \sim$ $10$ ps and   $100$ ps $\sim \hbar/U_{\rm max}\ll T_{\rm op}  \ll \hbar/U_{\rm min} \sim 10$ $\mu s$. These requirements can be satisfied in gate- and flux-controlled tuning of $E_{J1(2)}$ \cite{Hyart13, Aasen16}, as well as in gate-controlled tuning of $\Delta_{\rm ex}$. The operations should be performed much faster than the quasiparticle poisoning time, but since the external magnetic field is not required this does not pose additional constraints on the operation unless the poisoning time is  significantly shorter than the observed and predicted $T_{\rm pois}$ in the state-of-the-art devices \cite{Dima18,mannila2021superconductor,Dima21}.

\section{Conclusions}

We have shown that the combination of the proximity-induced superconductivity and the spontaneous TRS breaking allows the possibility to realize MZMs in band-inverted electron-hole bilayers in the absence of magnetic field.  We have studied the signatures of MZMs in superconductor/time-reversal symmetry broken insulator/superconductor Josephson junctions numerically using the full lattice model and analytically using the low-energy effective edge theory. We have shown that all the requirements for the observation of the $4\pi$ Josephson effect can be satisfied in this system. By modifying the setup so that the charging energy of the superconducting islands exceeds the temperature, it is possible detect the  Majorana fusion rules by utilizing tunable Josephson junctions and the gate-tunable energy gap opened by the  the spontaneous time-reversal symmetry breaking order parameter. Our estimates of the relevant energy scales indicate that all the requirements for robust Majorana fusion-rule detection can be satisfied in this system. 

\begin{acknowledgments}
We thank D. I. Pikulin for useful discussions and comments.
The work is supported by the Foundation for Polish Science through the IRA Programme
co-financed by EU within SG OP.  We acknowledge
the computational resources provided by
the Aalto Science-IT project and the access to the computing facilities of the Interdisciplinary Center of Modeling at the University of Warsaw, Grant No. G87-1164 and G78-13.
\end{acknowledgments}

\appendix

\section{Analytical solutions of effective low-energy theory for edge excitations \label{app:analytics}}

In this section, we derive analytical solutions of the effective low-energy edge Hamiltonian $H_e$, Eq.~(\ref{effham}), where $A_{{\rm eff}}(x)$ is the velocity of the helical edge states, and $\Delta_s(x)$ [$\Delta_{{\rm ex}}(x)$] determines the energy gap opened by the proximity induced superconductivity [time-reversal symmetry breaking order parameter] in the superconducting [normal] region of $x$.

\subsection{Solution of $H_e$ in the normal region}

In the normal regions, $\Delta_{{\rm ex}} > 0$ and $|\Delta_s(x)|=0$. The propagating solutions for $|E|>\Delta_{{\rm ex}}$ are
\begin{eqnarray}
\psi(x)&=&
b_1\begin{pmatrix}
1\\
i\delta\\
0\\
0\\
\end{pmatrix}e^{ik_{N}x}+b_2\begin{pmatrix}
-i\delta\\
1\\
0\\
0
\end{pmatrix}e^{-ik_{N}x} \nonumber \\ && +b_3\begin{pmatrix}
0\\
0\\
1\\
i\delta\\
\end{pmatrix}e^{-ik_{N}x}+b_4\begin{pmatrix}
0\\
0\\
-i\delta\\
1
\end{pmatrix}e^{ik_{N}x},
\label{wvfIIa}
\end{eqnarray}
where $\delta=\frac{\Delta_{{\rm ex}}}{\sqrt{E^2-\Delta_{{\rm ex}}^2}+E}$ and $k_{N}=\frac{\sqrt{E^2-\Delta_{{\rm ex}}^2}}{A^N_{{\rm eff}}}$, and the evanescent solutions for $|E|<\Delta_{{\rm ex}}$ are
\begin{eqnarray}
\psi(x)&=&p_1\begin{pmatrix}
e^{i\theta_1}\\
1\\
0\\
0
\end{pmatrix}e^{-\kappa_{N}x}+p_2\begin{pmatrix}
-e^{-i\theta_1}\\
1\\
0\\
0
\end{pmatrix}e^{\kappa_{N}x} \nonumber \\ && \hspace{-0.3cm} +p_3\begin{pmatrix}
0\\
0\\
e^{i\theta_1}\\
1
\end{pmatrix}e^{\kappa_{N}x}+p_4\begin{pmatrix}
0\\
0\\
-e^{-i\theta_1}\\
1
\end{pmatrix}e^{-\kappa_{N}x},
\label{wvfIIb}
\end{eqnarray}
where $e^{i\theta_1}=\frac{\sqrt{\Delta_{{\rm ex}}^2-E^2}-iE}{\Delta_{{\rm ex}}}$ and $\kappa_{N}=\frac{\sqrt{\Delta_{{\rm ex}}^2-E^2}}{A^N_{{\rm eff}}}$.

\subsection{Solution of $H_e$ in the superconducting region}

In the superconducting region, $\Delta_{{\rm ex}} = 0$ and $\Delta_s(x)=\Delta_0 e^{i \varphi}$. In our analytical calculations, we only need the decaying solutions for $|E|<\Delta_0$. The solutions decaying in $x \to -\infty$ and $x \to \infty$ have the form 
\begin{equation}
    \psi(x) = a_1\begin{pmatrix}
            e^{i\varphi}e^{-i\gamma}\\
            0 \\
            1 \\
            0
         \end{pmatrix} e^{\kappa_S x} + a_2\begin{pmatrix}
            0 \\
            e^{i\varphi}e^{i\gamma} \\
            0 \\
           1
         \end{pmatrix} e^{\kappa_S x},
\label{wvfI}        
\end{equation}
\begin{equation}
\psi(x) =c_1\begin{pmatrix}
            e^{i\varphi}e^{i\gamma}\\
            0 \\
            1 \\
            0
         \end{pmatrix} e^{-\kappa_S x} + c_2\begin{pmatrix}
            0 \\
            e^{i\varphi}e^{-i\gamma}\\
            0 \\
           1
         \end{pmatrix} e^{-\kappa_S x},
\label{wvfIII}         
\end{equation}
where  $e^{i\gamma}=\frac{E + i\sqrt{\Delta_0^2-E^2}}{\Delta_0}$ and  $\kappa_{S}=\frac {\sqrt{\Delta_0^2-E^2}}{A^S_{{\rm eff}}}$.

\subsection{Majorana zero mode at the interface of TRS broken and superconducting regions}

By combining the solutions (\ref{wvfIIb}) and (\ref{wvfIII}) with proper boundary conditions, it is easy to see that if the TRS broken insulator (superconductor) covers the region $x<0$ ($x>0$) there  is a MZM localized at $x=0$. Namely, there exist a zero-energy solution of the Hamiltonian (\ref{effham}) of the form
\begin{equation}
\psi(x)=  \begin{cases} \frac{1}{\cal{N}} \begin{pmatrix}
- e^{-i \pi/4} e^{i\varphi/2}   \\
  e^{-i \pi/4} e^{i \varphi/2}\\
 e^{i \pi/4} e^{-i \varphi/2}\\
 e^{i \pi/4} e^{-i \varphi/2}
\end{pmatrix} e^{\kappa_Nx}, & x<0 \\
\frac{1}{\cal{N}} \begin{pmatrix}
- e^{-i \pi/4} e^{i\varphi/2}  \\
e^{-i \pi/4} e^{i \varphi/2}\\
e^{i \pi/4} e^{-i \varphi/2}\\
 e^{i \pi/4} e^{-i \varphi/2}  
\end{pmatrix}e^{-\kappa_Sx}, & x>0
\end{cases}
\end{equation}
The corresponding field operator $\gamma$ in the second quantized form obeys $\gamma=\gamma^\dagger$, and by choosing the normalization constant ${\cal N}$ properly we obtain $\gamma^2=1$. Therefore, this solution satisfies the algebra of the MZMs.

\subsection{Energy spectrum of subgap states and  hybridization of MZMs across the TRS broken insulator in a Josephson junction}

In a Josephson junction the spatial profiles of $\Delta_{{\rm ex}}(x)$ and $\Delta_s(x)$ are 
\begin{equation}
\Delta_{{\rm ex}}(x) = \left\{
\begin{array}{lc}
\Delta_{{\rm ex}} & 0 \leq x \leq L \\
0 & \mathrm{elsewhere}
\end{array}
\right.
\end{equation}
and
\begin{equation}
\Delta_{s}(x) = \left\{
\begin{array}{lc}
\Delta_0 e^{-i\phi/2} & x < 0 \\
0 & 0 \leq x \leq L \\
\Delta_0e^{i\phi/2} & x > L
\end{array}
\right.\!.
\end{equation}

In the case $|E|>\Delta_{{\rm ex}}$, the solutions are of the form (\ref{wvfI}) for $x<0$, (\ref{wvfIIa}) for $0 \leq x \leq L$, and (\ref{wvfIII}) for $x>L$. Thus, the continuity of the wave function at $x=0$ and $x=L$ leads to constraints 
\begin{equation}
\begin{pmatrix}
a_1e^{-i\phi/2}e^{-i\gamma}\\
a_2e^{-i\phi/2}e^{i\gamma}\\
a_1\\
a_2
\end{pmatrix}=\begin{pmatrix}
b_1-i\delta b_2\\
i\delta b_1 +b_2\\
b_3-i\delta b_4\\
i\delta b_3 +b_4
\end{pmatrix},
\end{equation}
and
\begin{equation}
\begin{pmatrix}
b_1 e^{ik_{N}L} - i\delta b_2 e^{-ik_{N}L}\\
i\delta b_1 e^{ik_{N}L} + b_2 e^{-ik_{N}L}\\
b_3 e^{-ik_{N}L} - i\delta b_4 e^{ik_{N}L}\\
i\delta b_3 e^{-ik_{N}L} + b_4 e^{ik_{N}L} 
\end{pmatrix}=\begin{pmatrix}
c_1 e^{i\phi/2}e^{i\gamma}e^{-\kappa_S L}\\
c_2 e^{i\phi/2}e^{-i\gamma}e^{-\kappa_S L}\\
c_1 e^{-\kappa_S L}\\
c_2 e^{-\kappa_S L}
\end{pmatrix}.
\end{equation}
These equations can be written in matrix form as $M(a_1, a_2, b_1, b_2, b_3, b_4, c_1, c_2)^T=0$, which has non-trivial solutions only if $\det(M)=0$, so that this condition determines the allowed energies as a function of the parameters of the model $E(A^N_{{\rm eff}}, \Delta_{{\rm ex}}, \Delta_0, L, \phi)$.

In the case  $|E|<\Delta_{\rm ex}$, the solutions are of the form (\ref{wvfI}) for $x<0$, (\ref{wvfIIb}) for $0 \leq x \leq L$, and (\ref{wvfIII}) for $x>L$. Thus, the continuity of the wave function at $x=0$ and $x=L$ leads to constraints  
\begin{equation}
\begin{pmatrix}
a_1 e^{-i\phi/2}e^{-i\gamma}\\
a_2 e^{-i\phi/2}e^{i\gamma}\\
a_1\\
a_2
\end{pmatrix}=\begin{pmatrix}
p_1 e^{i\theta_1} -p_2 e^{-i\theta_1} \\
p_1+p_2\\
p_3 e^{i\theta_1} - p_4e^{-i\theta_1} \\
p_3 +p_4
\end{pmatrix},
\end{equation}
and
\begin{equation}
\begin{pmatrix}
p_1 e^{i\theta_1} e^{-\kappa_{N}L} - p_2 e^{-i\theta_1}e^{\kappa_{N}L}\\
p_1 e^{-\kappa_{N}L}+ p_2 e^{\kappa_{N}L}\\
p_3 e^{i\theta_1}e^{\kappa_{N}L} - p_4 e^{-i\theta_1} e^{-\kappa_{N}L} \\
p_3 e^{\kappa_{N}L} + p_4e^{-\kappa_{N}L}
\end{pmatrix}=\begin{pmatrix}
c_1 e^{i\phi/2}e^{i\gamma}e^{-\kappa_{S}L}\\
c_2 e^{i\phi/2}e^{-i\gamma}e^{-\kappa_S L}\\
c_1 e^{-\kappa_{S}L}\\
c_2 e^{-\kappa_{S}L}
\end{pmatrix},
\end{equation}
so that the allowed energies can be computed similarly as above. In the asymptotic limit $\kappa_N L \gg 1$, we obtain closed form solutions 
\begin{equation}
E=\pm 2 \frac{\Delta_0 \Delta_{{\rm ex}}}{\Delta_0+\Delta_{{\rm ex}}} e^{-\kappa_N L} \cos(\phi/2), \ \kappa_N \approx \Delta_{{\rm ex}}/A^N_{{\rm eff}}. \label{app:asymptotic}
\end{equation}
This equation describes the hybridization of the MZMs on across the TRS broken insulator in the superconductor/TRS broken insulator/superconductor Josephson junction in the asymptotic limit where the MZMs are weakly coupled. 

\begin{figure}[b]
    \centering
    \includegraphics[width=0.9\linewidth]{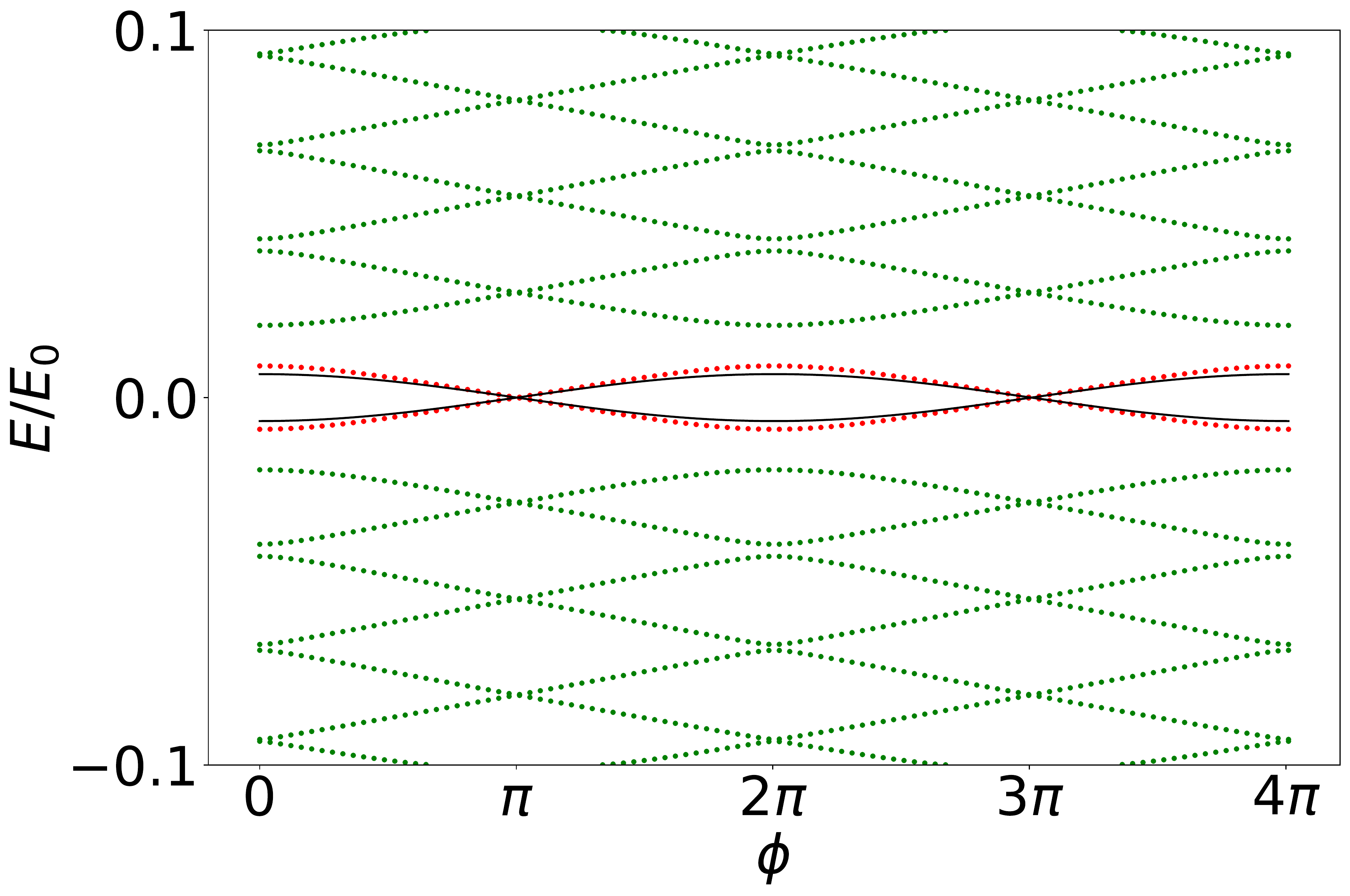}
    \caption{Energy spectrum for  $E_G^N=0.86E_0$, $\Delta_0=0.1E_0$ and $L_N=10d_0$ where the Majoranas appear at the interfaces of the system. The green (red) lines are the solutions with energy $E>\Delta_{{\rm ex}}$ ($E<\Delta_{{\rm exc}}$). The solid black line shows the analytic expression (\ref{app:asymptotic}).}
    \label{fig:spec}
\end{figure}

The complete subgap energy spectrum can be obtained by sweeping the energies $E$ from $-\Delta_0$ to $+\Delta_0$ and numerically finding the roots of determinants of the constraint matrices (see Fig.~\ref{fig:spec}).

\subsection{Hybridization of the MZMs across the superconducting region}

In a similar way we can also compute the hybridization of the MZMs on the opposite sides of a superconducting island. In this case, we assume that the spatial profiles of $\Delta_{{\rm ex}}(x)$ and $\Delta_s(x)$ are \begin{equation}
\Delta_{{\rm ex}}(x) = \left\{
\begin{array}{lc}
\Delta_{{\rm ex}}  & x < 0 \\
0 & 0 \leq x \leq L \\
\Delta_{{\rm ex}} & x > L
\end{array}
\right.
\end{equation}
and
\begin{equation}
\Delta_{s}(x) = \left\{
\begin{array}{lc}
\Delta_0  & 0 \leq x \leq L \\
0 & \mathrm{elsewhere}
\end{array}
\right.\!.
\end{equation}
In the asymptotic limit, the energies of the hybridized MZMs are
\begin{equation}
E=\pm 2 \frac{\Delta_0 \Delta_{{\rm ex}}}{\Delta_0+\Delta_{{\rm ex}}} e^{-\kappa_S L}, \ \kappa_S \approx \Delta_{{0}}/A^S_{{\rm eff}}.
\end{equation}

\section{Different approaches for calculating the Josephson effect and the system size dependence \label{app:Josephson}}

In this section, we briefly summarize the numerical approach for calculating the Josephson current-phase relationship with the hybrid kernel polynomial method (KPM) developed in Ref.~\onlinecite{irfan2019hybrid}. Moreover, we compare the hybrid KPM results to the ones obtained using exact diagonalization of small systems and the effective edge theory. Finally, we discuss the effects of system size on the magnitudes of the $4\pi$- and $2\pi$-periodic Josephson effect. 

\subsection{Hybrid kernel polynomial method for calculation of the supercurrent in a Josephson junction}

Although we are interested in the supercurrent $I$ at low temperatures, it is convenient to express it as 
\begin{align}
I = {\rm Tr}\left[\hat{I} f(\hat{H}_{BdG}) \right], \ \hat{I}=\frac{e}{\hbar}\frac{d \hat{H}_{BdG}}{d\phi},
\end{align}
where 
\begin{align}
 f(\hat{H}_{BdG})=\underset{k}{\sum} f(E_k) \ket{\psi_k} \bra{\psi_k},
\end{align}
$f(E)$ is the Fermi function and $\ket{\psi_k}$ are the eigenstates of the $\hat{H}_{BdG}$ with eigenenergies $E_k$. In the KPM method $\hat{H}_{BdG}$ needs to be scaled so that the spectrum $\{E_k\}$ is bounded to an interval $(-1,1)$ by choosing a suitable unit of energy. Then,  $f(E)$ can be expanded as \cite{KPM}
\begin{align}
    f(E)=\sum_{m=0}^{\infty} \alpha_m T_m(E),
\end{align}
where the Chebyshev's polynomials $T_m(x)=\cos(m \arccos{x})$ form a complete basis in $(-1,1)$ and they are orthogonal under the inner product
\begin{align}
    \langle f \cdot g \rangle = \int_{-1}^{1} \frac{f(x)g(x)}{\pi \sqrt{1-x^2}}dx,
\end{align}
so that the Chebyshev coefficients are given by $\alpha_m= \langle f(E) \cdot T_m(E) \rangle$. Thus, $f(\hat{H}_{BdG})$ can be written as
\begin{align}
    f(\hat{H}_{BdG})=\sum_{m=0}^{\infty} \alpha_m T_m(\hat{H}_{BdG}).
\end{align}
The series needs to be truncated to some order $M$, and in order to ensure stable convergence the method of Ref.~\cite{irfan2019hybrid}  utilizes Jackson kernel \cite{KPM}
\begin{align}
K_m = \frac{M-m+1}{M+1}\cos{\frac{\pi m}{M+1}} +\frac{1}{M+1}\frac{\sin{\frac{\pi m}{M+1}}}{\tan{\frac{\pi}{M+1}}}    
\end{align}
to modify the coefficients $\alpha_m$ to $\tilde{\alpha}_m=\alpha_m K_m$, so that
\begin{align}
    \tilde{f}(\hat{H}_{BdG}) = \sum_{m=0}^{M} \tilde{\alpha}_m T_m(\hat{H}_{BdG}).
    \label{eqn:Chebyshev_op_app}
\end{align}
The number of Chebyshev moments $M$ together with the choice of the kernel sets the energy resolution of the KPM approximation \cite{irfan2019hybrid, KPM}. The Fermi function changes rapidly around the Fermi level, and therefore in the hybrid KPM method \cite{irfan2019hybrid} a small subset of states $k\in A$ close to the Fermi energy is calculated exactly using sparse diagonalization method, so that 
\begin{equation}
    f(\hat{H}_{BdG})\approx \tilde{f}(\hat{H}_{BdG})+\sum_{k \in A} \big[f(E_k)-\tilde{f}(E_k)\big] \ket{\psi_k} \bra{\psi_k}.
\end{equation}
Thus the full expression for the total supercurrent is 
\begin{eqnarray}
\langle \hat{I} \rangle &\approx& \sum_{m=0}^{M} \tilde{\alpha}_m \bigg\{ {\rm Tr}\left[\hat{I} T_m(\hat{H}) \right]-  \sum_{k \in A}   T_m(E_k) \bra{\psi_k} \hat{I} \ket{\psi_k} \bigg\} \nonumber \\ && + \sum_{k \in A} f(E_k) \bra{\psi_k} \hat{I} \ket{\psi_k}. 
\label{app:total-supercurrent}
\end{eqnarray}
In the numerical calculations the superconducting phase difference is introduced through a Peierls substitution so that the expectation value of the current operator can be computed across a cut in the normal region separating the two superconductors \cite{irfan2019hybrid}.
Additionally we separate the total supercurrent into the components arising from the MZMs $I_M$ and the other Andreev levels. The first one gives the $4 \pi$ periodic contribution to the Josephson current and the latter one is responsible for the  $2\pi$ periodic Josephson effect.

\subsection{Comparison of the hybrid KPM method and exact diagonalization}

\begin{figure}[t]
    \centering
    \hspace{-0.4cm}
    \includegraphics[width=0.89\linewidth]{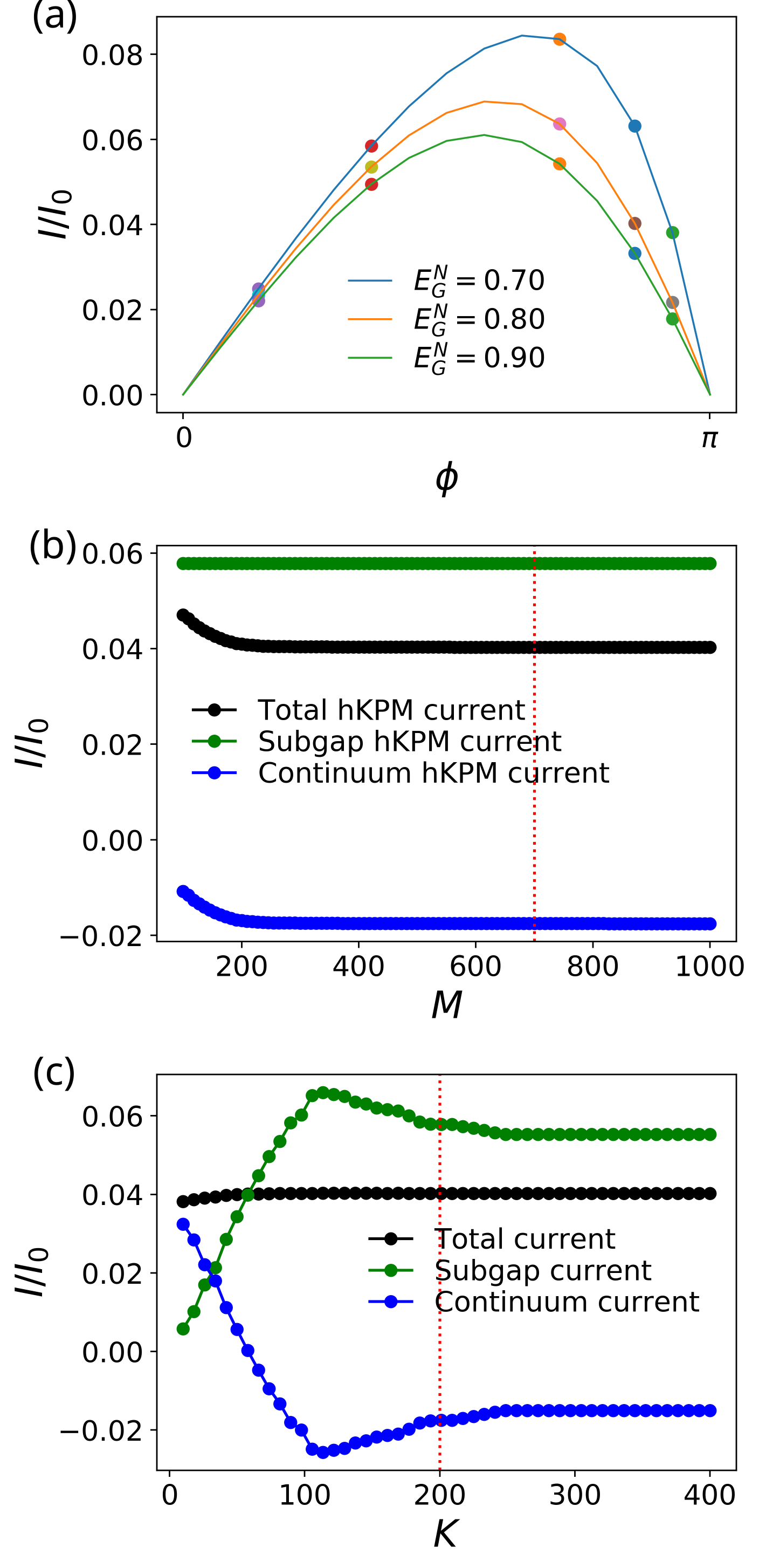}
    \caption{(a) Current-phase relationship for $L_N=10 d_0$, $W=60 d_0$, $L_S=15 d_0$, $\Delta_0=0.1E_0$, $E_G^N=:0.7 E_0, 0.8E_0, 0.9 E_0$ and $E_G^S=1.3E_0$.  The solid lines show the current calculated using hybrid kernel polynomial method and the circles show the exact results for particular values of $\phi$. Here, the number of moments is $M=700$ and the maximum number of computed subgap states is $K=200$. (b) Convergence of the subgap, continuum and the total  current with increasing $M$ for $K=200$  for $\phi=2.69$ and $E_G^N=0.8E_0$. (c) Same with increasing $K$ for $M=700$.
    }
    \label{fig:exact&tot}
\end{figure}

In the case of small systems we can compare the hybrid KPM results to the supercurrent obtained using exact diagonalization. For this purpose we consider the device geometry shown in Fig.~\ref{fig:josephson_junction}. We find that for sufficiently large number of moments $M=700$ and the maximum number of computed subgap states $K=200$ the hybrid KPM results are in good agreement with the exact results (see  Fig.~\ref{fig:exact&tot}). In all our hybrid KPM calculations we have used $M \geq 700$ and $K \geq 200$.

\subsection{Comparison of the hybrid KPM and the effective edge theory results}

\begin{figure}[h]
    \centering
    \includegraphics[width=0.8\linewidth]{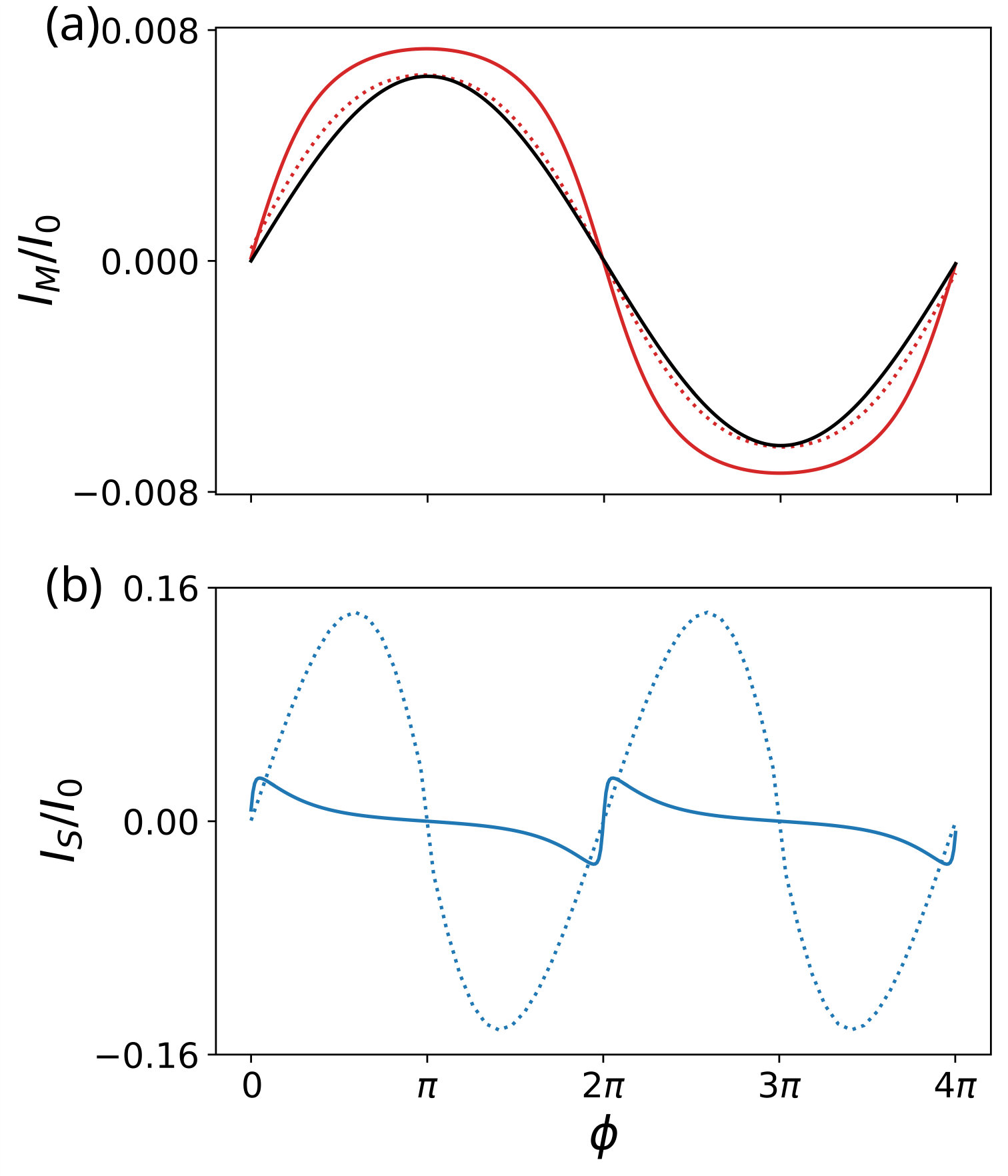}
    \caption{ 
    (a) The $4\pi$-periodic supercurrent $I_M(\phi)$ originating from the MZMs. The low-energy effective theory (solid red line) agrees well with the numerical results obtained from the lattice model (dashed red line). The results are also in good agreement with the analytical expression (black line) given by Eq.~(\ref{I_M-analytic_app}).  
    (b) The $2\pi$-periodic supercurrent $I_S(\phi)$ originating from the rest of the states. The $I_S(\phi)$ of the full lattice model (dashed blue line)  deviates significantly from the results of the low-energy effective edge theory (solid blue line)  because there exists many high-energy Andreev levels with significant dispersion as a function of $\phi$.
    The corresponding energy spectra for the lattice model  and continuum model are shown in Figs.~\ref{fig:spectrum}(b) and \ref{fig:spec}, respectively.
    The model parameters are $E_G^N=0.86E_0$, $L_N=10d_0$, $W=150 d_0$, $E_G^S=1.3 E_0$ and $L_S=15 d_0$ and $\Delta_0=0.1E_0$.   
    }
    \label{fig:en_current_comp}
\end{figure}

We have also compared the hybrid KPM results to the results obtained using the low-energy effective edge theory. As can be seen in Fig.~\ref{fig:en_current_comp} the $4\pi$-periodic component $I_M(\phi)$ originating from the MZMs is well captured by the effective edge theory. In the asymptotic limit $k_NL \gg 1$ we obtain from Eq.~(\ref{app:asymptotic})
\begin{equation}
 I_M(\phi)= 2  \frac{e}{\hbar}\frac{\Delta_0 \Delta_{{\rm ex}}}{\Delta_0+\Delta_{{\rm ex}}} e^{-\Delta_{{\rm ex}} L_N/A^N_{{\rm eff}}} \frac{{\cal P}_{12}+{\cal P}_{34}}{2}  \sin (\phi/2). \label{I_M-analytic_app}
 \end{equation}
We find that this analytic expressions accurately describes the parametric dependencies of the $4\pi$-periodic current in the regime of weakly broken TRS. (In all figures we plot $I_M(\phi)$ for ${\cal P}_{12}=1$ and ${\cal P}_{34}=1$.) On the other hand, the low-energy effective edge theory strongly underestimates the magnitude of the $2\pi$-periodic component, because in the full lattice model there exists more Andreev levels with significant dispersion as a function of $\phi$ than in the case of the effective edge theory.

\subsection{System size dependence of the Josephson effects}

\begin{figure}
    \centering
    \includegraphics[width=\linewidth]{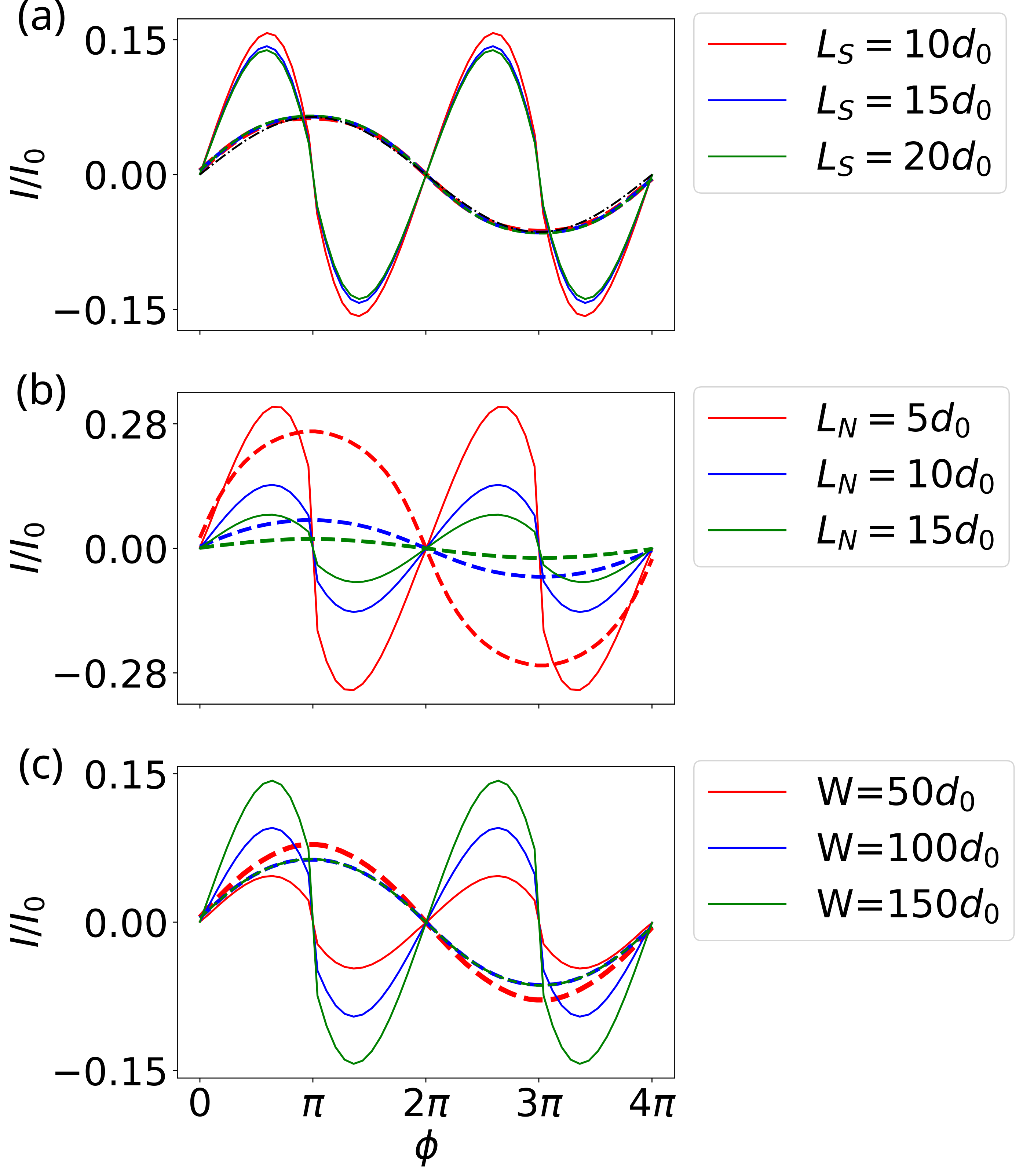}
    \caption{(a) The $4\pi$-periodic supercurrent $I_M(\phi)$ originating from the MZMs (dashed, magnified by 10 times) and the $2\pi$-periodic current $I_S(\phi)$ (solid) originating from the rest of the states for $L_N=10 d_0$, $W=150 d_0$ and $L_S=:10, 15, 20 d_0$. The analytic approximation Eq.~(\ref{I_M-analytic_app}) is shown with dashdot black line.  (b) Same for $W=150 d_0$, $L_S=15 d_0$ and  $L_N=:5, 10, 15 d_0$. (c) Same for $L_S=15 d_0$,  $L_N= 10 d_0$ and $W=: 50,100, 150 d_0$
    The other parameters are $\Delta_0=0.1E_0$, $E_G^N=0.86E_0$ and $E_G^S=1.3E_0$. 
    }
    \label{fig:size_vary}
\end{figure}

We have also calculated the dependence of the $2\pi$ and $4\pi$ periodic Josephson effects on the system size (see Fig.~\ref{fig:size_vary}). 
We find that $L_s$ does not affect the results significantly as long as $L_s>10 d_0$. As expected increasing $L_N$ decreases both the $2\pi$ and $4\pi$ periodic Josephson effects. On the other hand, decreasing $W$ reduces the magnitude of the $2\pi$ Josephson effect but it does not influence significantly the $4 \pi$ periodic Josephson effect. 

\section{Fusion of Majorana zero modes \label{app:fusion}}

In the fusion setup shown in Fig.~\ref{fig:fusion_setup} the four MZMs, described by the operators $\gamma_i=\gamma_i^\dag$ ($i=A,B,C,D$) satisfying $\{\gamma_i,\gamma_j\}=2\delta_{ij}$, are related to  fermonic operators 
\begin{align}
 f^\dag_{AB}=\frac 12 (\gamma_A+i\gamma_B), \quad f^\dag_{CD}=\frac 12 (\gamma_C+i\gamma_D).
\end{align}
The occupation operators of these fermions are given by
\begin{eqnarray}
 \hat{N}_{AB}&=&f_{AB}^\dagger f_{AB}=\frac{1-i\gamma_A\gamma_B}{2}, \nonumber \\  
 \hat{N}_{CD}&=&f_{CD}^\dagger f_{CD}=\frac{1-i\gamma_C\gamma_D}{2}.
\end{eqnarray}
In the case of even total parity, the two possible states of the system are
\begin{equation}
|0_{AB}, 0_{CD}\rangle, \quad |1_{AB}, 1_{CD}\rangle=f_{AB}^\dag f_{CD}^\dag |0_{AB}, 0_{CD}\rangle
\end{equation}
satisfying
\begin{eqnarray}
 \hat{N}_{AB} |N_{AB} N_{CD}\rangle &=& N_{AB} |N_{AB} N_{CD}\rangle, \nonumber\\ 
  \hat{N}_{CD} |N_{AB} N_{CD}\rangle &=& N_{CD} |N_{AB} N_{CD}\rangle.
\end{eqnarray}
Alternatively, we can form the fermion operators, number operators and states as
\begin{eqnarray}
 f_{AD}^\dag =\frac 12 (\gamma_A+i\gamma_D), \quad f_{BC}^\dag=\frac 12 (\gamma_B+i\gamma_C),
\end{eqnarray}
\begin{eqnarray}
 \hat{N}_{AD}&=&f_{AD}^\dagger f_{AD}=\frac{1-i\gamma_A\gamma_D}{2}, \nonumber \\  \hat{N}_{BC}&=&f_{BC}^\dagger f_{BC}=\frac{1-i\gamma_B\gamma_C}{2}
\end{eqnarray}
and
\begin{equation}
|0_{AD}, 0_{BC}\rangle, \quad |1_{AD}, 1_{BC}\rangle=f_{AD}^\dag f_{BC}^\dag |0_{AD}, 0_{BC}\rangle.
\end{equation}

The two sets of fermionic operators are related as
\begin{eqnarray}
 f_{AD}^\dagger &=& \frac 12 (f_{AB}+f_{AB}^\dagger-f_{CD}+f_{CD}^\dagger), \nonumber \\
 f_{BC}^\dagger &=& \frac i2 (f_{AB}-f_{AB}^\dagger +f_{CD}+f_{CD}^\dagger).
\end{eqnarray}
Thus, by straightforward algebra we find that the basis states are related as
\begin{eqnarray}
 |0_{AD}0_{BC}\rangle&=&\frac 1{\sqrt 2} \big(|0_{AB}0_{CD}\rangle+|1_{AB}1_{CD}\rangle\big), \nonumber \\
  |1_{AD}1_{BC}\rangle&=&\frac i{\sqrt 2} \big(|1_{AB}1_{CD}\rangle-|0_{AB}0_{CD}\rangle\big).
\end{eqnarray}
The important consequence of this simple algebra is that if one initializes the system into a state with well-defined fermion numbers $N_{AD}$ and $N_{BC}$, and then measures the fermion number $N_{AB}$ (or $N_{CD}$), the outcome will be $0$ and $1$ with equal probabilities. This is a manifestation of the fusion rule of the MZMs
\begin{equation}
\sigma \times \sigma = I+\psi,
\end{equation}
which states that the MZMs can coalesce into identity $I$ or a fermion $\psi$.

\bibliography{bibliography}

\end{document}